\documentclass[reprint,pre,twocolumn,showpacs,superscriptaddress]{revtex4-2}
\usepackage{amsmath,amsthm,amssymb,amscd,float,graphicx,bbold}
\usepackage[shortlabels]{enumitem}
\setlist[enumerate,1]{1.,itemsep=2pt,parsep=0pt,topsep=3pt,left=0pt}
\usepackage[utf8]{inputenc}
\usepackage[T1]{fontenc}
\usepackage{lmodern}
\usepackage[colorlinks,linkcolor=blue,citecolor=blue,urlcolor=blue]{hyperref}
\newcommand{\al}{\alpha}
\newcommand{\be}{\beta}
\newcommand{\de}{\delta}

\newcommand{\vep}{\varepsilon}
\newcommand{\ga}{\gamma}
\newcommand{\ka}{\kappa}
\newcommand{\la}{\lambda}

\newcommand{\si}{\sigma}

\newcommand{\ze}{\zeta}
\newcommand{\De}{\Delta}

\newcommand{\La}{\Lambda}

\newcommand{\bbs}{\mathbf{s}}

\newcommand{\bv}{\mathbf{v}}
\newcommand{\bV}{\mathbf{V}}

\newcommand{\bga}{{\boldsymbol{\ga}}}

\newcommand{\tS}{\widetilde{S}}

\newcommand{\CC}{{\mathbb C}}

\newcommand{\cE}{{\mathcal E}}

\newcommand{\cH}{{\mathcal H}}

\newcommand{\pd}{\partial}

\newcommand{\id}{1\hspace{-.25em}{\rm I}}
\newcommand{\noi}{\noindent}

\newcommand{\ket}[1]{|#1\rangle}

\newcommand{\mss}{\kern 1pt}

\renewcommand{\le}{\leqslant}
\renewcommand{\ge}{\geqslant}
\newcommand{\tends}[1]{\bbuildrel{\hbox to 2em{\rightarrowfill}}_{#1}^{}}

\newcommand{\csch}{\operatorname{csch}}

\newcommand{\Li}{\operatorname{Li}}
\newcommand{\tr}{\operatorname{tr}}

\newcommand{\iu}{\mathrm i}
\newcommand{\diff}{\mathrm{d}}

\newcommand{\su}{\mathrm{su}}

\newcommand{\gl}{\mathrm{gl}}
\newcommand{\en}{\enspace}

\newcommand{\Int}[1]{\,\mathop{\!#1}\limits^{\lower1ex\hbox{$\scriptstyle\circ$}}{}}

\theoremstyle{remark}
\newtheorem{remark}{Remark}

\newcommand{\bsv}{\mathbf s}

\def\clap#1{\hbox to 0pt{\hss#1\hss}}

\begin{document}

\title{Thermodynamics and criticality of supersymmetric\\ spin chains of Haldane--Shastry type}

\author{Federico Finkel}\email{ffinkel@ucm.es}
\author{Artemio González-López}\email[Corresponding author. Email address: ]{artemio@ucm.es}

\affiliation{Depto.~de Física Teórica, Facultad de Ciencias Físicas,\\
  Universidad Complutense de Madrid,\\
  Plaza de las Ciencias 1, 28040 Madrid, SPAIN}
\date{\today}
\begin{abstract}
  We analyze the thermodynamics and criticality properties of four families of $\su(m|n)$
  supersymmetric spin chains of Haldane--Shastry (HS) type, related to both the $A_{N-1}$ and the
  $BC_N$ classical root systems. Using a known formula expressing the thermodynamic free energy
  per spin of these models in terms of the Perron (largest in modulus) eigenvalue of a suitable
  inhomogeneous transfer matrix, we prove a general result relating the $\su(kp|kq)$ free energy
  with arbitrary $k=1,2,\dots$ to the $\su(p|q)$ free energy. In this way we are able to evaluate
  the thermodynamic free energy per spin of several infinite families of supersymmetric HS-type
  chains, and study their thermodynamics. In particular, we show that in all cases the specific
  heat at constant volume features a single marked Schottky peak, which in some cases can be
  heuristically explained by approximating the model with a suitable multi-level system with
  equally spaced energies. We also study the critical behavior of the models under consideration,
  showing that the low-temperature behavior of their thermodynamic free energy per spin is the
  same as that of a $(1+1)$-dimensional conformal field theory with central charge $c=m+n/2-1$.
  However, using a motif-based description of the spectrum we prove that only the three families
  of $\su(1|n)$ chains of type $A_{N-1}$ and the $\su(m|n)$ HS chain of $BC_N$ type with $m=1,2,3$
  (when the sign $\vep_B$ in the Hamiltonian takes the value $-1$ in the latter case) are truly
  critical.

\end{abstract}

\maketitle

\section{Introduction}\label{sec.intro}

In this paper we shall deal with a large class of $\su(m|n)$ supersymmetric spin chains of
Haldane--Shastry type whose spectrum can be concisely described in terms of bond vectors and
(supersymmetric) Haldane motifs~\cite{HHTBP92,BGHP93}. More precisely, these chains consist of $N$
sites occupied by either a boson with $m$ internal degrees of freedom or a fermion with $n$
internal degrees of freedom. We shall consider the following two types of models:

\smallskip \noi I) The supersymmetric versions of the Haldane--Shastry
(HS)~\cite{Ha88,Sh88,Ha93,BB06}, Polychronakos--Frahm (PF)~\cite{Po93,Fr93,Po94,HB00} and
Frahm--Inozemtsev (FI)~\cite{FI94,BBH10,BFGR10,FGLR18} chains of $A_{N-1}$ type. Their Hamiltonians can
be jointly expressed as~\footnote{In what follows sums over Latin indices will implicitly range
  from $1$ to $N$, unless otherwise stated.}
\begin{equation}\label{HANm1}
  H = \sum_{i<j}J_{ij}(1-S_{ij})\,,
\end{equation}
where $S_{ij}$ is an $\su(m|n)$-supersymmetric spin permutation operator and the couplings
$J_{ij}$ are given by
\begin{alignat}{2}
  J_{ij}&=\frac{J/N^2}{2\sin^2(\xi_i-\xi_j)}, &&\xi_k=\frac{k\pi}N
                                                     \quad (\text{HS})\,,\label{HS}\\
  J_{ij}&=\frac{J/N}{(\xi_i-\xi_j)^2}\,, &&H_N(\xi_k)=0
                                                \quad (\text{PF})\,,\label{PF}\\
  J_{ij}&=\frac{J/N^2}{2\sinh^2(\xi_i-\xi_j)}\,,\quad &&L_N^{\al-1}(e^{2\xi_k})=0
                                                        \quad (\text{FI})\label{FI}
\end{alignat}
(see Appendix~\ref{app.Sij} for a review of the definition of the operators $S_{ij}$).
Here~$J\ne0$ is a real constant setting the energy scale, $H_N$ denotes the Hermite polynomial of
degree $N$ and $L_N^{\al-1}$ is a generalized Laguerre polynomial of degree $N$ with parameter
$\al>0$. It is apparent from the previous formulas that the HS chain can be regarded as a circular
chain with equally spaced sites and spin-spin interactions inversely proportional to the square of
the chord distance. On the other hand, the PF and FI chains are linear chains with sites $\xi_k$
defined in Eqs.~\eqref{PF}-\eqref{FI} and respectively rational or hyperbolic inverse square
interactions.

\smallskip \noi II) The supersymmetric version of the Haldane--Shastry chain of $BC_N$
type (HS-B)~\cite{BPS95,EFGR05,CFGR22}, with Hamiltonian
\begin{subequations}\label{Bchain}
\begin{multline}
  H=\frac J{4N^2}\sum_{i<j}\bigg(\frac{1-S_{ij}}{\sin^{2} (\xi_i-\xi_j)}
  +\frac{1-\widetilde{S}_{ij}}{\sin^{2}(\xi_i+\xi_j)}\bigg)\\
  +\frac{J}{8N^2}\sum_i\left(\frac{b_1}{\sin^{2} \xi_i} + \frac{b_2}{\cos^2
      \xi_i}\right)\big(1-S_i\big),
\end{multline}
where
\begin{equation}
  P_N^{b_1-1,b_2-1}(\cos 2\xi_k)=0.
\end{equation}
\end{subequations}
Here $J\ne0$, $P_N^{b_1-1,b_2-1}$ is a Jacobi polynomial of degree $N$ and parameters $b_{1,2}>0$,
$S_i$ is an $\su(m|n)$-supersymmetric spin reversal operator (see Appendix~\ref{app.Sij} for a
review of its precise definition) and we have used the standard
abbreviation~$\tS_{ij}:=S_{ij}S_iS_j=S_iS_jS_{ij}$. The HS-B chain can be regarded as the
\emph{open} version of the HS chain, whose sites (in general not uniformly spaced)
$z_j:=e^{2\iu\xi_j}$ lie on the upper unit circle. Each spin interacts both with the remaining
spins and their reflections with respect to the real axis, the interaction strength being
inversely proportional in both cases to the square of the chord distance.

The type-$A_{N-1}$ chains~\eqref{HANm1}--\eqref{FI} are remarkable instances of many-body
integrable models~(see, e.g., \cite{Ka92,BGHP93,HH93}). Moreover, both the HS and the PF chains
are exactly invariant under the Yangian~$Y[\gl(m|n)]$ for any number of
spins~\cite{Ha94,Hi95,BUW99,HB00}, and the same is conjectured to be true for the FI
chain~\cite{BBH10}. Exploiting this symmetry, it has been shown that the energy spectrum of the
three type-$A_{N-1}$ chains is the same as that of an inhomogeneous vertex model with $N+1$
vertices~\cite{BBH10}. More precisely, let us take the sets of bosonic and fermionic degrees of
freedom respectively as
\begin{equation}\label{BFA}
  B=\{1,\dots,m\}\,,\qquad F=\{m+1,\dots,m+n\}.
\end{equation}
Labeling the vertices by the integers~$0,\dots,N$, any two consecutive vertices $i$ and $i+1$
(with $i=0,\dots,N$) are joined by a \emph{bond} $s_i$ taking values in $\{1,\dots,m+n\}$. The
model's configurations are the set of \emph{bond vectors}
\[
  \bbs:=(s_1,\dots,s_N)\in\{1,\dots,m+n\}^N,
\]
whose energy is defined by
\begin{equation}
  \label{bondE}
 E(\bbs)=J\sum_{i=1}^{N-1}\de(s_i,s_{i+1})\cE(x_i)\,,\quad x_i:=i/N\,.
\end{equation}
In the latter formula the \emph{dispersion function} $\cE(x)$ is given by
\begin{equation}\label{disp}
  \cE(x)=\begin{cases}
    x(1-x),\quad &(\text{HS})\\
  x, &(\text{PF})\\
  x(x+\ga_N),&(\text{FI})
\end{cases}
\end{equation}
with $\ga_N:=(\al-1)/N$, while the function $\de:(B\cup F)^2\to\{0,1\}$ is defined by
\begin{equation}\label{delta}
  \de(\mu,\nu)=\begin{cases}
    0,\en \mu<\nu\quad\text{or}\quad \mu=\nu\in B\\
  1,\en \mu>\nu\quad\text{or}\quad \mu=\nu\in F.
\end{cases}
\end{equation}
It is shown in Ref.~\cite{BBH10} that the complete spectrum of the three $A_{N-1}$-type
chains~\eqref{HANm1}--\eqref{FI}, with the correct degeneracy of each energy level, is obtained
from Eqs.~\eqref{bondE}--\eqref{delta} as the bond vector $\bbs$ runs over $\{1,\dots,m+n\}^N$. In
fact, the vectors $(\de(s_1,s_2),\dots,\de(s_{N-1},s_N))\in\{0,1\}^{N-1}$ are essentially
supersymmetric Haldane \emph{motifs}~\cite{Ha94}.

It was recently shown that the spectrum of the HS-B chain~\eqref{Bchain} also admits a description
in terms of a variant of supersymmetric motifs~\cite{CFGR22}. More precisely, let us define the
integers $m_{\vep_B}$ and $n_{\vep_F}$ by
\[
  m_{\vep_B}=\frac12\big(m+\vep_B\pi(m)\big),\qquad n_{\vep_F}=\frac12\big(n+\vep_F\pi(n)\big),
\]
where $\vep_B$ and $\vep_F$ are the two signs appearing in the definition of the spin reversal
operators $S_i$ (see Appendix~\ref{app.Sij}) and $\pi(k)=\frac12\,\big(1-(-1)^k\big)$ is the
parity of the integer $k$, and take the sets of bosonic and fermionic degrees of freedom
respectively as
\begin{equation}\label{BFB}
  F=\{m_{\vep_B}+1,\dots,m_{\vep_B}+n\},\qquad B=\{1,\dots,m+n\}\setminus F.
\end{equation}
In this case we must add an additional vertex $N+2$ and a corresponding last bond $s_{N+1}$ taking
the \emph{fixed} (half-integer) value
\begin{equation}\label{sNp1}
  s_{N+1}=m_{\vep_B}+n_{\vep_F}+\frac12\equiv s^*.
\end{equation}
The energy spectrum can then be generated through the formula
\begin{equation}\label{EbsvB}
  E(\bsv)=J\sum_{i=1}^{N}\de(s_i,s_{i+1})\cE(x_i),\qquad x_i=i/N\,,
\end{equation}
with $\de$ given again by Eq.~\eqref{delta} and dispersion function
\begin{equation}
  \label{dispB}
  \cE(x)=x\,\bigg(\ga_N+1-\frac{x}2\bigg),
  \quad\ga_N:=\frac1{2N}(b_1+b_2-1)\,.
\end{equation}

As shown in Refs.~\cite{EFG12,FGLR18,FG22jstat}, equations~\eqref{bondE}-\eqref{EbsvB} for the
energy spectrum are the key ingredient in the evaluation of the free energy of the
chains~\eqref{HANm1}--\eqref{Bchain} in the thermodynamic limit. Indeed, using the latter
equations it is straightforward to rewrite the partition function $Z_N$ as the trace of a product
of matrices of order $m+n$. More precisely, we have
\begin{equation}\label{ZN}
  Z_N=\tr\big[A(x_0)A(x_1)\cdots A(x_{N-1})B\big]\,,
\end{equation}
where the \emph{transfer matrix} $A(x)$ has entries
\begin{equation}
  \label{Amat}
  A_{\mu\nu}(x)=e^{-\be J\cE(x)\de(\mu,\nu)}\,,\quad 1\le\mu,\nu\le m+n,
\end{equation}
$\be:=1/T$ is the inverse temperature (taking Boltzmann's constant $k_{\mathrm B}$ as unity), and
the dispersion relation~$\cE(x)$ is given by Eqs.~\eqref{disp}-\eqref{dispB}. The
$(m+n)\times(m+n)$ matrix $B$ is the identity for the three chains of type $A_{N-1}$, while for
the HS-B chain its entries are given by
\[
  B_{\mu\nu}=e^{-\be J(\ga_N+1/2)\de(\mu,s^*)}\,,\quad 1\le\mu,\nu\le m\,.
\]
Since $A_{\mu\nu}(x)>0$ for all $x\in[0,1]$, the classical Perron theorem~\cite{Pe07,GK00}
guarantees that $A(x)$ has a positive simple eigenvalue $\la_1(x)$ which is strictly greater than
the modulus of any other eigenvalue. From this fact and Eq.~\eqref{ZN} it readily
follows~\cite{FGLR18,FG22jstat} that in the thermodynamic limit $N\to\infty$ the free energy per
spin $f(T)$ of the chains~\eqref{HANm1}--\eqref{FI} and~\eqref{Bchain} can be expressed
as~\footnote{It is understood that the parameter $\ga_N$ appearing in the dispersion relation of
  the FI and HS-B chains must be replaced by its $N\to\infty$ limit $\ga$ in Eq.~\eqref{f}.}
\begin{equation}
  \label{f}
  f(T)=-T\int_0^1\ln\la_1(x)\,\diff x\,.
\end{equation}

It is clear from the previous discussion that in order to obtain a closed-form expression for the
thermodynamic free energy of the chains~\eqref{HANm1}--\eqref{FI} and~\eqref{Bchain}, and thereby
derive their main thermodynamic functions, it suffices to compute the Perron eigenvalue~$\la_1(x)$
of the transfer matrix~\eqref{Amat}. This has been done so far in the non-supersymmetric
$\su(m|0)$ (purely bosonic) or $\su(0|n)$ (purely fermionic) cases~\cite{FG22pre} for arbitrary
$m$ and $n$. The explicit calculation of the Perron eigenvalue $\la_1(x)$ of the transfer matrix
$A(x)$ in the genuinely supersymmetric case $mn\ne0$ is much harder due to the smaller degree of
symmetry of the transfer matrix. Indeed, in the supersymmetric case the Perron eigenvalue
$\la_1(x)$ is only explicitly known for $m,n\le 2$~\cite{FGLR18,FG22jstat}.

One of the main aims of this work is to compute in closed form the Perron eigenvalue
$\la_1(x)\equiv\la_1^{(m|n)}(x)$ of the $\su(m|n)$ supersymmetric chains of Haldane--Shastry
type~\eqref{HANm1}--\eqref{Bchain}, and through it the main thermodynamic functions, for higher
values of $m$ and $n$. The key idea in this respect is to find a simple, explicit relation between
the Perron eigenvalue $\la^{(p|q)}_1$ and $\la_1^{(kp|kq)}$ for $k=1,2,\dots$, which will make it
possible to extend the previously known results for low values of $p$ and $q$ to arbitrary
multiples thereof. In addition, we shall find new explicitly solvable cases with $m,n\le 3$ (and
their corresponding multiples). As a byproduct of our analysis, we shall derive a simple
asymptotic formula for the $\su(m|n)$ partition function and main thermodynamic functions in the
limit of large $m$ and $n$. In particular, since the asymptotic expression for the energy and
specific heat turn out to be independent of $m$ and $n$, this shows that these functions become
independent of $m$ and $n$ for large values of these parameters.

The intimate connection between spin chains of HS type and conformal field theories (CFTs) in
$(1+1)$ dimensions has been known for a long time. Indeed, already in 1992 Haldane and
collaborators~\cite{HHTBP92} proved that the low energy excitations of the $\su(0|m)$ HS chain are
governed by the $\su(m)$ WZNW model at level $1$ (see also~\cite{Sc94,BS96}). Later on, Hikami and
Basu-Mallick~\cite{HB00} showed that the partition function of the $\su(m|n)$ supersymmetric PF
chain is related to the highest-weight character of the $\su(m|n)_1$ model when the number of
spins tends to infinity. Although this result might naively be construed as implying that the
$\su(m|n)$ PF chain is critical (i.e., described at low energies by an effective
$(1+1)$-dimensional CFT in the thermodynamic limit), it should be taken into account that a
genuinely critical quantum system must possess a ground state with finite degeneracy and admit low
energy excitations above this ground state with a linear energy-momentum relation. It was later
shown in Ref.~\cite{BBS08} that this is only the case for the $\su(m|n)$ HS chain with $m=0$ or
$1$. This result was shown to hold also for the $\su(0|n)$ PF, FI and HS-B chains in
Ref.~\cite{FG22pre}, although the latter chain was shown to be critical also in the $\su(2|0)$ and
$\su(3|0)$ cases (with $\vep_B=-1$ in the latter case). The second aim of this work is to
determine in what cases the supersymmetric PF, FI and HS-B chains are truly critical, and to
compute the central charge of their corresponding CFTs. To determine the central charge, we shall
use the well-known result which posits that at low temperatures the free energy per unit length of
a CFT behaves as
\begin{equation}\label{fc}
  f(T)=f(0)-\frac{\pi c T^2}{6v_F}+o(T^2),
\end{equation}
where $c$ is the central charge of the Virasoro algebra and $v_F$ is the Fermi
velocity~\cite{BCN86,Af86}. Although in general the Perron eigenvalue $\la_1(x)$ is not known
explicitly, we shall find a novel closed-form expression for the characteristic polynomial of the
matrix $A(x)$ valid for arbitrary values of $m$ and $n$. From this expression we shall determine
the low temperature limit of $f(T)$ for all $m$ and $n$ through equation~\eqref{f}, and thus
evaluate the central charge $c$ in all the critical cases. Using the motif-based description of
the spectrum discussed above, we shall then study the ground state degeneracy and the existence of
low energy excitations with linear energy-momentum relation for all nonzero values of $m$ and $n$.
In this way we shall prove that the PF and FI chains are only critical when $m=1$ (and $\ga>0$ for
the FI chain), while the HS-B chain is also critical for $m=2$ and $m=3$ (with $\vep_B=-1$ in the
latter case).

The paper's organization is as follows. In Section~\ref{sec.fe} we prove our main result on the
Perron eigenvalue of the $\su(m|n)$ transfer matrix $A(x)$ when $m$ and $n$ have a common
multiple, and apply it to derive exact formulas for the free energy per spin of several infinite
families of supersymmetric spin chains of HS type. With the help of these formulas, we study the
thermodynamics of the latter models in Section~\ref{sec.thermo}. In Section~\ref{sec.CB} we
analyze the low temperature behavior of the $\su(m|n)$ free energy per spin for arbitrary $m$ and
$n$, showing that it behaves as the free energy of a $(1+1)$-dimensional CFT with central charge
$c=m+n/2-1$. Using the motif.based description of the spectrum outlined above, we then determine
all the values of $m$ and $n$ for which $\su(m|n)$ chains of HS type are truly critical. In
Section~\ref{sec.conc} we present our conclusions, and outline a few topics for future research
suggested by the present work. The paper ends with four technical appendices in which we present
the definition of the supersymmetric permutation and spin reversal operators, establish a property
of the eigenvalues of supersymmetric transfer matrices used in Section~\ref{sec.fe}, derive an
asymptotic formula for an integral used in the study of the low-temperature behavior of the free
energy, and summarize a few properties of the dilogarithm function needed in the latter section.

\section{Perron eigenvalue and free energy}\label{sec.fe}

In this section we shall derive our main results on the free energy per spin of $\su(m|n)$
supersymmetric chains of HS type. To begin with, we shall exploit the known symmetry of the
partition function of these models under the exchange of bosons and fermions~\cite{BBHS07} to
restrict ourselves, without loss of generality, to the case of positive $J$. Indeed, for the
$A_{N-1}$-type chains~\eqref{HANm1}--\eqref{FI} the partition function of the $\su(m|n)$ and
$\su(n|m)$ chains with couplings $\pm J$ are related by
\[
  Z_N^{(m|n)}(T;J)=e^{-J\be E_0}Z_N^{(n|m)}(T;-J),
\]
where
\[
  E_0=\sum_{i=1}^{N-1}\cE(x_i)=\begin{cases}
    \frac{N^2-1}{6N},& \text{HS},\\[1mm]
    \frac12(N-1),& \text{PF},\\[1mm]
    \frac{N-1}{6N}\big[(3\ga_N+2)N-1\big],& \text{FI}.
  \end{cases}
\]
In terms of the thermodynamic free energy, the previous relations become
\begin{equation}\label{fJmJ}
  f^{(m|n)}(T;J)=J\cE_0+f^{(n|m)}(T;-J),
\end{equation}
where
\[
  \cE_0=\lim_{N\to\infty}\frac{E_0}N=\begin{cases}
    \frac16,& \text{HS},\\[1mm]
    \frac12,& \text{PF},\\[1mm]
    \frac{\ga}{2}+\frac13,& \text{FI},
  \end{cases}
\]
with $\ga=\lim_{N\to\infty}\ga_N$. The type-$BC_N$ chain~\eqref{Bchain} is also symmetric under
boson-fermion exchange, provided that we  replace $(\vep_B,\vep_F)$ by $(-\vep_F,-\vep_B)$. More
precisely, in this case we have~\cite{FG22jstat}
\[
  Z_N^{(m,\vep_B|n,\vep_F)}(T;J)=e^{-J\be E_0}Z_N^{(n,-\vep_F|m,-\vep_B)}(T;-J),
\]
with
\[
  E_0=\frac{N-1}{12N}\big[2(3\ga_N+2)N+1\big],\qquad \text{HS-B}.
\]
Since the thermodynamic free energy is independent of the signs
$\vep_B$ and $\vep_F$, this is easily seen to imply Eq.~\eqref{fJmJ} with
\[
  \cE_0=\frac{\ga}{2}+\frac13,\qquad \text{HS-B},
\]
as for the FI chain of type $A_{N-1}$.

We shall next prove one of the main results of the paper, which relates the thermodynamic free
energies per spin $f^{(p|q)}(T)$ and $f^{(kp|kq)}(T)$ of two $\su(p|q)$ and $\su(kp|kq)$
supersymmetric chains of HS type, where $k=1,2,\dots$. More precisely, we shall show that
\begin{align}
  \la_1^{(kp|kq)}(x;J)&=\la_1^{(k|0)}(x;J)\la_1^{(p|q)}(x;J/k)\notag\\
                      &=\frac{1-e^{-\be J\cE(x)}}{1-e^{-\be J\cE(x)/k}}\,\la_1^{(p|q)}(x;J/k),
                          \label{la1pqk}
\end{align}
which by Eq.~\eqref{f} implies the remarkable relation
\begin{equation}
  \label{fpqkpq}
  f^{(kp|kq)}(T;J)=f^{(k|0)}(T;J)+f^{(p|q)}(T;J/k).
\end{equation}

As shown in Ref.~\cite{FG22pre}, a necessary and sufficient condition for an eigenvalue of a
positive matrix to be its Perron eigenvalue is that it possesses a \emph{positive eigenvector}
(i.e., an eigenvector whose components are all strictly positive). We start by expressing the
components of the (unique, up to normalization) positive eigenvector~\footnote{In what follows, we
  shall often suppress the argument $x\in(0,1)$ for the sake of conciseness.} $\bv$ of the Perron
eigenvalue $\la_1$ for the transfer matrix $A\equiv A^{(p|q)}$. To this end, we first take
advantage of the invariance of the eigenvalues of the transfer matrix ---and, in particular, of
its Perron eigenvalue $\la_1$--- under rearrangements of its diagonal elements
(cf.~Appendix~\ref{app.Adiag}) to write the latter matrix as~\footnote{In what follows we shall
  implicitly assume that $mn\ne0$, referring the reader to Ref.~\cite{FG22pre} for the purely
  bosonic or fermionic cases $m=0$.}
\begin{equation}\label{A}
  A=
  \begin{pmatrix}
    1&1 &\cdots &\cdots &\cdots &1\\
    a&\ddots &\ddots & & &\vdots\\
    \vdots &\ddots &1 &1 & &\vdots\\
    \vdots & &a &a &\ddots &\vdots\\
    \vdots & & &\ddots &\ddots &1\\
    a &\cdots &\cdots &\cdots &a &a
  \end{pmatrix}\equiv A^{(p|q)}[a],
\end{equation}
with
\begin{equation}\label{adef}
  a=e^{-J\be\cE(x)},
\end{equation}
both in the $A_{N-1}$ and the $BC_N$ cases. In other words, all the
elements above (resp.~below) the main diagonal of the matrix~\eqref{A} are $1$'s (resp.~$a$'s),
while the main diagonal is made up of $p$ $1$'s followed by $q$ $a$'s. Writing
$\bv=(v_0,\dots,v_{p+q-1})$ and setting $v_0=1$ without loss of generality (since $\bv$ is
positive), the eigenvector $\bv$ satisfies the system of equations
\begin{alignat}{3}
  \label{sumvila1}
  1+\sum_{i=1}^{p+q-1}v_i&=\la_1\\
  a+a\sum_{i=1}^{j-1}v_i+\sum_{i=j}^{p+q-1}v_i&=\la_1 v_j, &&j=1,\dots,p-1\notag\\
  a+a\sum_{i=1}^{p+l-1}v_i+\sum_{i=p+l}^{p+q-1}v_i&=\la_1 v_{p+l-1},\qquad &&l=1,\dots,q.\notag
\end{alignat}
Combining the first and the last equations we obtain
\begin{equation}\label{la1v}
  v_{p+q-1}=a,\qquad \la_1=a+1+\sum_{i=1}^{p+q-2}v_i.
\end{equation}
From the first set of equations we deduce that
\begin{equation*}
  (a-1)v_j=\la_1(v_{j+1}-v_j)\implies v_{j+1}=\left(1+\frac{a-1}{\la_1}\right)v_j,
\end{equation*}
for $j=0,\dots,p-2$, which taking into account that $v_0=1$ is easily solved to yield
\begin{equation}\label{vj}
  v_j=\left(1+\frac{a-1}{\la_1}\right)^j,\qquad j=0,\dots,p-1.
\end{equation}
Likewise, form the second set of equations satisfied by $\bv$ we obtain
\begin{multline*}
  (a-1)v_{p+l}=\la_1(v_{p+l}-v_{p+l-1})\\
  \implies v_{p+l-1}=\left(1-\frac{a-1}{\la_1}\right)v_{p+l}
\end{multline*}
for $l=1,\dots,q-1$, whence (taking into account that $v_{p+q-1}=a$)
\begin{equation}\label{vppl}
  v_{p+l}=a\left(1-\frac{a-1}{\la_1}\right)^{q-l-1},\qquad l=0,\dots,q-1.
\end{equation}
In particular, since $\la_1$ is the Perron eigenvalue of $A$ and $a>0$ by Eq.~\eqref{la1v} we have
\begin{equation}\label{la1ge1}
  \la_1>a+1>1,
\end{equation}
which is consistent with the positivity of the components of $\bv$ by
Eqs.~\eqref{vj}-\eqref{vppl}. Substituting Eqs.~\eqref{vj}-\eqref{vppl} into Eq.~\eqref{la1v} and
operating we obtain the following algebraic equation satisfied by $\la_1$:
\begin{equation}
  \label{eigveq}
  \left(1+\frac{a-1}{\la_1}\right)^p-a\left(1-\frac{a-1}{\la_1}\right)^q=0.
\end{equation}
In fact, the previous calculation is clearly valid for \emph{any} nonzero eigenvalue $\la$ of the
transfer matrix $A$. This shows that the characteristic polynomial $P_A(\la)=\det(A-\la)$ of the
latter matrix is given by~\footnote{Note that the left-hand side of Eq.~\eqref{eigveq} vanishes
  identically when $a=1$, and is thus divisible by $1-a$.}
\[
  P_A(\la)=\frac{(-1)^{p+q}}{1-a}\left[\la^q(\la+a-1)^p-a\la^p(\la-a+1)^q\right].
\]

We are now ready to prove the fundamental relation~\eqref{la1pqk}. To this end, let us first of
all set
\[
  b=a^{1/k}=e^{-\be J\cE(x)/k},
\]
and rearrange the diagonal elements of the transfer matrix $A^{(kp|kq)}\equiv A^{(kp|kq)}[a]$ of
the $\su(kp|kq)$ chain as
\[
  \big(\underbrace{1,\dots,1}_{p}\,,\underbrace{b^k,\dots,b^k}_q\,,
  \dots,\underbrace{1,\dots,1}_{p}\,,\underbrace{b^k,\dots,b^k}_q\,\big),
\]
i.e., as $k$ sequences of $p$ $1$'s followed by $q$ $b^k$'s. Note that, by the result in
Appendix~\ref{app.Adiag}, this rearrangement does not alter the eigenvalues of the transfer
matrix. We shall establish Eq.\eqref{la1pqk} by showing that
\begin{equation}\label{La1}
  \La_1:=\frac{1-b^k}{1-b}\,\la_1[b]
\end{equation}
is an eigenvalue of the transfer matrix $A^{(kp|kq)}$ with eigenvector
\[
  \bV:=\big(\bv,b\bv,\dots,b^{k-1}\bv\big),
\]
where $\la_1[b]$ denotes the Perron eigenvalue of the matrix~$A^{(p|q)}[b]$ in Eq.~\eqref{A} and
$\bv\equiv\bv[b]$ is its positive eigenvector with components given by Eq.~\eqref{vj}-\eqref{vppl}
(with $a$ replaced by $b$). Since $\bV$ is clearly a \emph{positive} vector, Eq.~\eqref{La1}
implies that $\La_1$ is the Perron eigenvalue of the transfer matrix $A^{(kp|kq)}$ and
establishes~\eqref{la1pqk}.

To prove our claim, consider the submatrix $A_l^{(kp|kq)}$ of $A^{(kp|kq)}$ consisting of the
rows $l(p+q)+1,\dots,(l+1)(p+q)$ (with $l=0,\dots,k-1$) of the latter matrix. This submatrix can
be schematically represented as
\[
  A_l^{(kp|kq)}=\left(
  \begin{array}{ccc|ccc|ccc}
    b^k & \cdots & b^k & && & 1 & \cdots & 1\\
    \vdots &  & \vdots & &A^{(p|q)}[b^k]& & \vdots & & \vdots\\
    b^k& \cdots & b^k & & &&1 & \cdots & 1
  \end{array}
  \right),
\]
where $A^{(p|q)}[b^k]$ occupies the columns
\[
  l(p+q)+1,\dots,(l+1)(p+q)
\]
of $A_l^{(kp|kq)}$. To show that $\bV$ is an eigenvector of $A^{(kp|kq)}$ with eigenvalue $\La_1$,
it suffices to check that
\[
  A_l^{(kp|kq)}\cdot \bV=b^l\La_1\bv,\qquad l=0,\dots,k-1.
\]
Taking into account the definition of $\bV$, these equations can be more explicitly written as
\[
  b^{-l}\left(b^k\sum_{i=0}^{l-1}b^i+\sum_{i=l+1}^{k-1}b^i\right)\sum_{j=0}^{p+q-1}v_j+A^{(p|q)}[b^k]\cdot\bv
  =\La_1\bv.
\]
Performing the geometric sums, and using Eq.~\eqref{sumvila1} and the definition of $\La_1$, the
latter $k$ equations are easily seen to reduce to the single equation
\begin{equation}\label{eigveqLa1}
  \frac{b^{k}-b}{b-1}+\la_1[b]^{-1}A^{(p|q)}[b^k]\cdot\bv=\frac{b^{k}-1}{b-1}\,\bv\,.
\end{equation}
To verify this equation, note that the components of the vector $A^{(p|q)}[b^k]\cdot\bv$ are given by
\begin{subequations}\label{Avcomps}
  \begin{equation}\label{firstcomp}
    (A^{(p|q)}[b^k]\cdot\bv)_j=b^k\sum_{i=0}^{j-1}v_i+\sum_{i=j}^{p+q-1}v_i
  \end{equation}
  for $j=1,\dots,p-1$, and
  \begin{equation}\label{lastcomp}
    (A^{(p|q)}[b^k]\cdot\bv)_{p-1+j}=b^k\sum_{i=0}^{p-1+j}v_i+\sum_{i=p+j}^{p+q-1}v_i
  \end{equation}
\end{subequations}
for $j=1,\dots,q- 1$. Using Eqs.~\eqref{vj}-\eqref{vppl} for the components of $\bv$ and
performing the geometric sums in Eq.~\eqref{firstcomp} we obtain
\begin{align*}
  &\la_1^{-1}(A^{(p|q)}[b^k]\cdot\bv)_j=\frac{b^k}{b-1}\left[\left(1+\frac{b-1}{\la_1}\right)^j-1\right]\\
  &\quad+\frac{1}{b-1}\left[\left(1+\frac{b-1}{\la_1}\right)^{p}
    -\left(1+\frac{b-1}{\la_1}\right)^{j}\right]\\
  &\quad-\frac{b}{b-1}\left[\left(1-\frac{b-1}{\la_1}\right)^{q}
    -1\right]=\frac{b-b^k}{b-1}+\frac{b^{k}-1}{b-1}\,v_j,
\end{align*}
where we have taken into account the eigenvalue equation~\eqref{eigveq} satisfied by $\la_1$. The
previous calculation establishes the validity of the first $p$ components of
Eq.~\eqref{eigveqLa1}. Likewise, Eq.~\eqref{lastcomp} can be expanded as
\begin{align*}
  &\la_1^{-1}(A^{(p|q)}[b^k]\cdot\bv)_{p-1+j}=
    \frac{b^k}{b-1}\left[\left(1+\frac{b-1}{\la_1}\right)^p-1\right]\\
  &\quad-\frac{b^{k+1}}{b-1}\left[\left(1-\frac{b-1}{\la_1}\right)^{q}
    -\left(1+\frac{b-1}{\la_1}\right)^{q-j}\right]\\
  &\quad-\frac{b}{b-1}\left[\left(1-\frac{b-1}{\la_1}\right)^{q-j}
    -1\right]
  \\
  &=\frac{b-b^k}{b-1}+\frac{b^{k}-1}{b-1}\,v_{p-1+j},
\end{align*}
where we have again used the eigenvalue equation~\eqref{eigveq}. Hence the last $q$ components of
Eq.~\eqref{eigveqLa1} also hold, which completes the proof of our claim that~\eqref{La1} is the
Perron eigenvalue of the transfer matrix $A^{(kp|kq)}$.

Combining the fundamental relation~\eqref{La1} just proved with Eq.~\eqref{f} we arrive at the
identity:
\begin{multline}
  \label{fkpkq}
  f^{(kp|kq)}(T)=-T\int_0^1\diff x\ln\left(\frac{1-e^{-J\be\cE(x)}}{1-e^{-J\be\cE(x)/k}}\right)\\
  -T\int_0^1\diff x\,\ln\left(\la_1^{(p|q)}\bigl[e^{-J\be\cE(x)/k}\bigr]\right),
\end{multline}
which can be used to evaluate $f^{(kp|kq)}$ for any positive integer $k$ when $f^{(p|q)}$ is
known. To begin with, the explicit formulas for the Perron eigenvalues of the $\su(1|1)$,
$\su(2|1)$, and $\su(1|2)$ chains from Ref.~\cite{FGLR18}, namely
\begin{align*}
  \la_1^{(1|1)}[a]&=1+a,\\
  \la_1^{(1|2)}[a]&=\frac12\left(1+2a+\sqrt{1+8a}\,\right),\\
  \la_1^{(2|1)}[a]&=1+\frac12\left(a+\sqrt{a(a+8)}\,\right)
\end{align*}
with $a$ given by Eq.~\eqref{adef}, yield closed-form expressions for the thermodynamic free
energies per spin of the $\su(k|k)$, $\su(k|2k)$ and $\su(2k|k)$ HS-type chains with arbitrary
$k=1,2,\dots$. We have also computed in closed form the Perron eigenvalue of the $\su(p|q)$ chains
with $p$ or $q$ equal to three (and $p\ne q$), which are given by
\begin{align*}
  \la^{(1|3)}[a]&=\frac13(1+3a)+\frac13\De_{13}^{1/3}+\frac13(1+15a)\De_{13}^{-{1/3}},\\
  \la^{(3|1)}[a]&=\frac13(a+3)+\frac13a^{1/3}\De_{31}^{1/3}+\frac13a^{2/3}(a+15)\De_{31}^{-{1/3}},\\
  \la^{(2|3)}[a]&=\frac13(2+3a)+\frac13\De_{23}^{1/3}+\frac13(1+24a)\De_{23}^{-{1/3}},\\
   \la^{(3|2)}[a]&=\frac13(3+2a)+\frac13a^{1/3}\De_{32}^{1/3}+\frac13a^{2/3}(a+24)\De_{32}^{-{1/3}},
\end{align*}
with
\begin{align*}
  \De_{13}&=27a^2+36a+1+3(1-a)\sqrt{3a(27a+1)}\,,\\
  \De_{31}&=a^2+36a+27+3(1-a)\sqrt{3(a+27)}\,,\\
  \De_{23}&=54a^2+72a-1+6(1-a)\sqrt{3a(27a-2)}\,,\\
  \De_{32}&=-a^2+72a+54+6(1-a)\sqrt{3(27-2a)}\,.
\end{align*}
Note that $\De_{pq}$, and hence $\la^{(p|q)}[a]$, is clearly positive for $0\le a\le1$.
\begin{remark}
  A simple consistency check of the relation~\eqref{fpqkpq} is obtained by computing
  $f^{(klp|klq)}$ both from $f^{(p|q)}$ and from $f^{(kp|kq)}$. In other words, we should have
  \begin{align*}
    &f^{(klp|klq)}(T;J)=f^{(kl|0)}(T;J)+f^{(p|q)}(T;J/kl)\\
                      &\quad=f^{(l|0)}(T;J)+f^{(kp|kq)}(T;J/l)\\
                      &\quad=f^{(l|0)}(T;J)+f^{(k|0)}(T;J/l)+f^{(p|q)}(T;J/kl),
  \end{align*}
  i.e.,
  \[
    f^{(kl|0)}(T;J)=f^{(l|0)}(T;J)+f^{(k|0)}(T;J/l),
  \]
  or equivalently
  \[
    \la_1^{(kl|0)}(T;J)=\la_1^{(l|0)}(T;J)\la_1^{(k|0)}(T;J/l).
  \]
  In fact, the latter equality is an immediate consequence of the identity
  \[
    \la_1^{(m|0)}(T;J)=\frac{1-a}{1-a^{1/m}},
  \]
  with $a$ given by Eq.~\eqref{adef}.\qed
\end{remark}

  We shall close this section by listing a few general properties of the thermodynamic free energy
  per spin that easily follow from the previous discussion. (In what follows we shall assume, as
  above, that we are dealing with the genuinely supersymmetric case $mn\ne0$.)
\begin{enumerate}
\item As $T\to0+$ we have $a=e^{-J\be\cE(x)}\to0$, and hence $\la_1(x)\to 1$ by the eigenvalue
  equation~\eqref{eigveq}. It follows that $f(0)=0$.
\item By Eq.~\eqref{la1ge1}, $f(T)<0$ for all $T>0$. In fact, for $T>0$ we have
  \[
    f(T)<-T\int_0^1\diff x\ln\left(1+e^{-J\be\cE(x)}\right)=f^{(1|1)}(T).
  \]
\item Likewise, when $T\to\infty$ we have $a\to1$, and therefore the transfer matrix $A^{m|n}(x)$
  tends to the matrix with all elements equal to $1$. Since the Perron eigenvalue of this matrix
  is $m+n$, it follows that
  \begin{equation}\label{fTinf}
    f(T)\underset{T\to\infty}\sim -T\ln(m+n),
  \end{equation}
  which is consistent with the known property
  \[
    Z_N\underset{T\to\infty}\sim (m+n)^N.
  \]
\item It is also straightforward to extract the behavior of $f^{(kp|kq)}$ as $k\to\infty$ (with
  $p,q$ fixed) from Eq.~\eqref{La1} for the Perron eigenvalue~$\La_1\equiv\la^{(kp|kq)}_1$.
  Indeed, in this limit we have
  \[
    \la_1[b]=\la_1[a^{1/k}]\underset{k\to\infty}\to\la_1[1]=p+q
  \]
  and
  \[
    \frac{1-b^k}{1-b}=\frac{1-e^{-J\be\cE(x)}}{1-e^{-J\be\cE(x)/k}}
    \underset{k\to\infty}\sim\frac{k(1-e^{-J\be\cE(x)})}{J\be\cE(x)},
  \]
  whence
  \[
    \la_1^{(kp|kq)}\underset{k\to\infty}\sim k(p+q)\frac{1-e^{-J\be\cE(x)}}{J\be\cE(x)}.
  \]
  Equation~\eqref{f} then yields the asymptotic formula
  \begin{multline*}
    f^{(kp|kq)}(T)\underset{k\to\infty}\sim-\ln\bigl(k(p+q)\bigr)T\\-
    T\int_0^1\diff x\,\ln\left(\frac{1-e^{-J\be\cE(x)}}{J\be\cE(x)}\right).
  \end{multline*}
\end{enumerate}
\begin{figure}[t]
  \centering
  \includegraphics[width=\columnwidth]{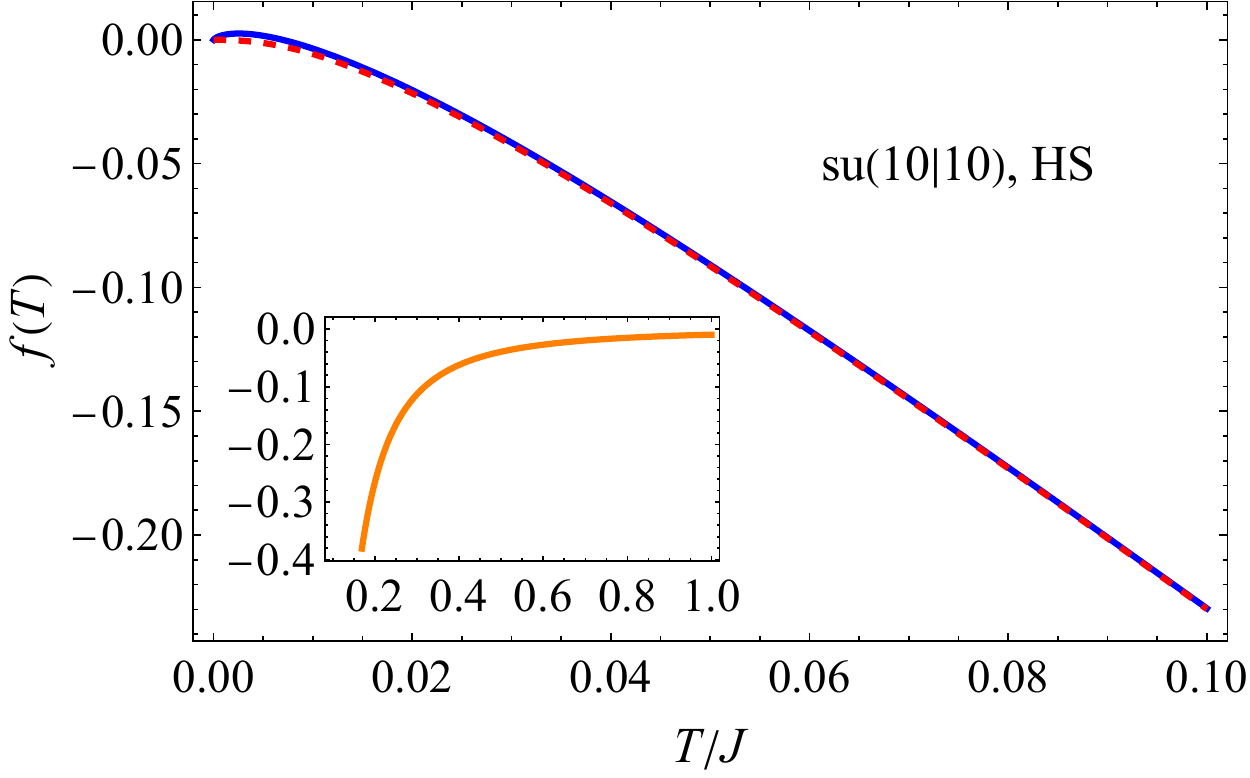}
  \caption{Thermodynamic free energy per spin of the $\su(10|10)$ HS chain compared to its
    asymptotic approximation~\eqref{fapp}. Inset: relative error of the latter approximation of
    $f^{(10|10)}(T)$ multiplied by $10^3$.}
  \label{fig.fapp}
\end{figure}
The asymptotic behavior of the $\su(kp|kq)$ free energy per spin as $k\to\infty$ strongly suggests
that when $m$ and $n$ are both large but otherwise arbitrary the more general asymptotic equality
\begin{multline}\label{fapp}
  f^{(m|n)}(T)\underset{m,n\to\infty}\sim-\ln\bigl(m+n)T\\-
  T\int_0^1\diff x\,\ln\left(\frac{1-e^{-J\be\cE(x)}}{J\be\cE(x)}\right)
\end{multline}
should hold. In fact, this relation can be proved by noting that the eigenvalue
equation~\eqref{eigveq} can be written as
\begin{equation}\label{mueq}
  \mu^{m/n}-a^{1/n}(2-\mu)=0,
\end{equation}
where
\[
  \mu:=1-\frac{1-a}{\la_1}\le 1.
\]
As $n\to\infty$ we have $a^{1/n}=e^{-J\be\cE(x)/n}\to 1$ for $T\gtrsim J/n$, and Eq.~\eqref{mueq}
tends to
\[
  \mu^{m/n}=2-\mu.
\]
Since $0<\mu\le 1$, and the previous equation has the root $\mu=1$, we can set $\mu=1-\vep$ with
$\vep\ge0$ small. Inserting this expression into Eq.~\eqref{mueq} we obtain
\[
  \mu^{m/n}\sim 1-\frac{m}n\vep=a^{1/n}(1+\vep),
\]
whence
\[
  \vep\sim\frac{1-a^{1/n}}{a^{1/n}+m/n}\sim\frac{1-e^{-\be J\cE(x)/n}}{m/n+1}\sim
  \frac{J\be\cE(x)}{m+n}.
\]
Since
\[
  \la_1=\frac{1-a}{1-\mu}\sim\frac{1-a}\vep=(m+n)\frac{1-e^{-J\be\cE(x)}}{J\be\cE(x)},
\]
substituting this expression into Eq.~\eqref{f} yields the asymptotic estimate~\eqref{fapp}. Note
that the second term in Eq.~\eqref{fapp} is independent of $m$ and $n$, and tends to zero as
$T\to\infty$, while the first term accounts for the correct asymptotic behavior~\eqref{fTinf} as
$T\to\infty$. The asymptotic approximation~\eqref{fapp} is quite accurate for temperatures
$T\gtrsim J/n$ even for relatively low values of $m$ and $n$ (see, e.g., Fig.~\ref{fig.fapp} for
the case $m=n=10$).

\section{Thermodynamics}\label{sec.thermo}

From the explicit expression of the free energy per spin of the supersymmetric spin chains of HS
type listed at the end of the previous section it is straightforward to compute all the relevant
thermodynamic functions using the standard formulas
\[
  u=\frac{\pd}{\pd\be}(\be f),\qquad s=\be(u-f),\qquad c_V=-\be^2\frac{\pd u}{\pd\be}\,.
\]
Here $u$, $s$, and $c_V$ respectively denote the energy, entropy and specific heat (at constant
volume) per spin. The corresponding formulas are particularly simple for the self-dual $\su(k|k)$
chain, whose free energy per spin can be written as
\begin{align}
  f^{(k|k)}(T)&= -T\int_0^1\diff x\ln\Bigl(\big(1-e^{-\be\cE(x)}\big)
                \coth\bigl(\tfrac{\be\cE(x)}{2k}\bigr)\Bigr)\notag\\
              &=\frac{\cE_0}2
                -T\int_0^1\diff x\ln\Bigl(2\coth\bigl(\tfrac{\be\cE(x)}{2k}\bigr)
                \sinh\bigl(\tfrac{\be\cE(x)}2\bigr)\Bigr).
    \label{f11}
\end{align}
Here, as in the rest of this Section, we have set $J=1$, which amounts to measuring the
temperature in units of $J$. From the previous expression for the free energy we easily obtain
\begin{align*}
  u^{(k|k)}(T)&=\frac{\cE_0}2+\int_0^1\diff x\,\cE(x)
                \left[\frac1k\csch\bigl(\tfrac{\be\cE(x)}k\bigr)\right.\\
              &\hphantom{=\frac{\cE_0}2+\int_0^1\diff x\cE(x)\bigg[\frac1k}
                -\left.\frac12\coth\bigl(\tfrac{\be\cE(x)}2\bigr)\right],\\
  s^{(k|k)}(T)&=\int_0^1\diff x\left[\vphantom{\frac1k}
                \ln\Bigl(2\coth\bigl(\tfrac{\be\cE(x)}{2k}\bigr)
                \sinh\bigl(\tfrac{\be\cE(x)}2\bigr)\Bigr)\right.\\
              &+\left.\be\cE(x)\left(
                \frac1k\csch\bigl(\tfrac{\be\cE(x)}k\bigr)
                -\frac12\coth\bigl(\tfrac{\be\cE(x)}2\bigr)\right)\right],\\
  c_V^{(k|k)}(T)&=\be^2\int_0^1\diff x\,\cE^2(x)\left[\frac1{k^2}
                  \coth\bigl(\tfrac{\be\cE(x)}k\bigr)\csch\bigl(\tfrac{\be\cE(x)}k\bigr)\right.\\
              &\hphantom{=\be^2\int_0^1\diff x\,\cE^2(x)\bigg[\frac1{k^2}}
                \left.-\frac14\csch^2\bigl(\tfrac{\be\cE(x)}2\bigr)\right].
\end{align*}%
\begin{figure}[t]
  \includegraphics[height=.3\columnwidth]{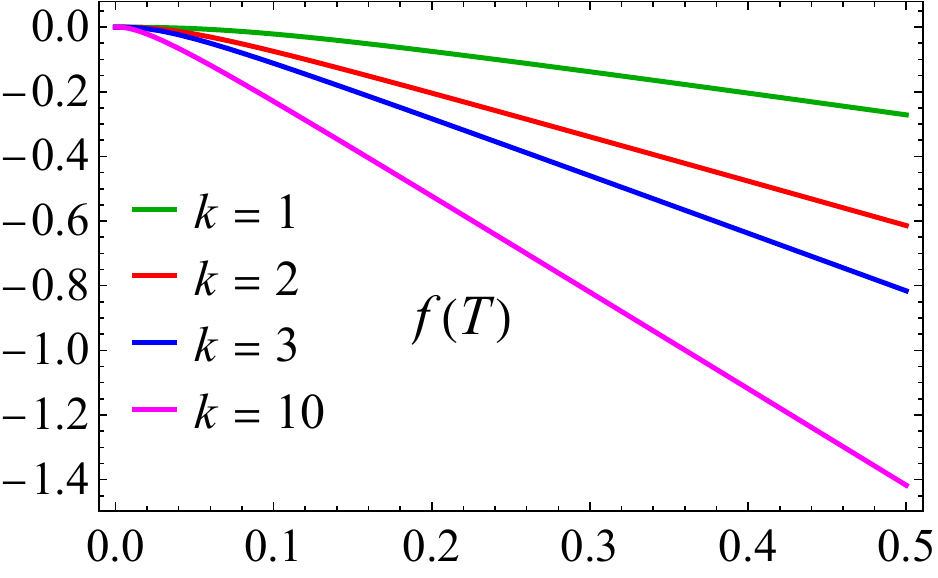}\hfill
  \includegraphics[height=.3\columnwidth]{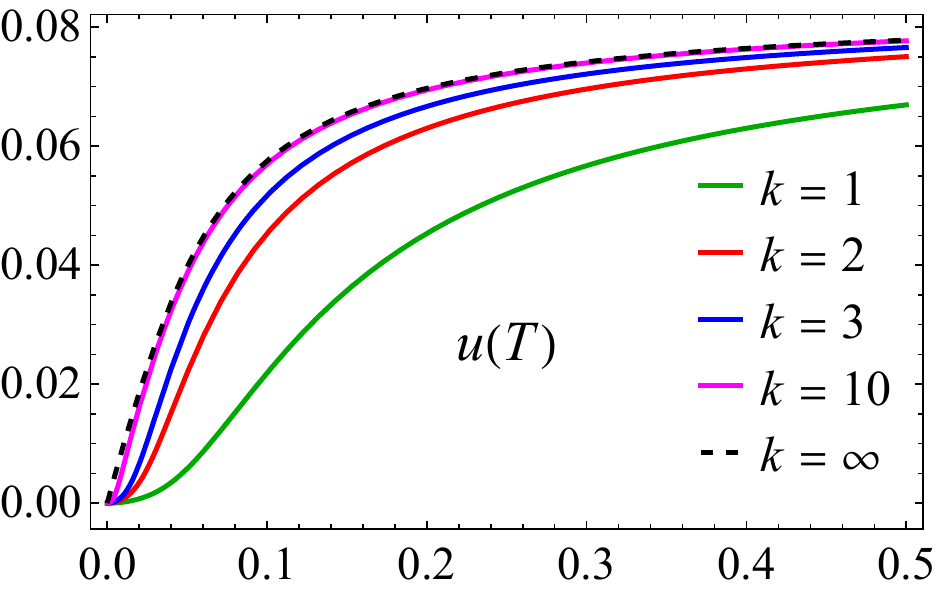}\\[3pt]
  \null\en\kern3pt\includegraphics[height=.3\columnwidth]{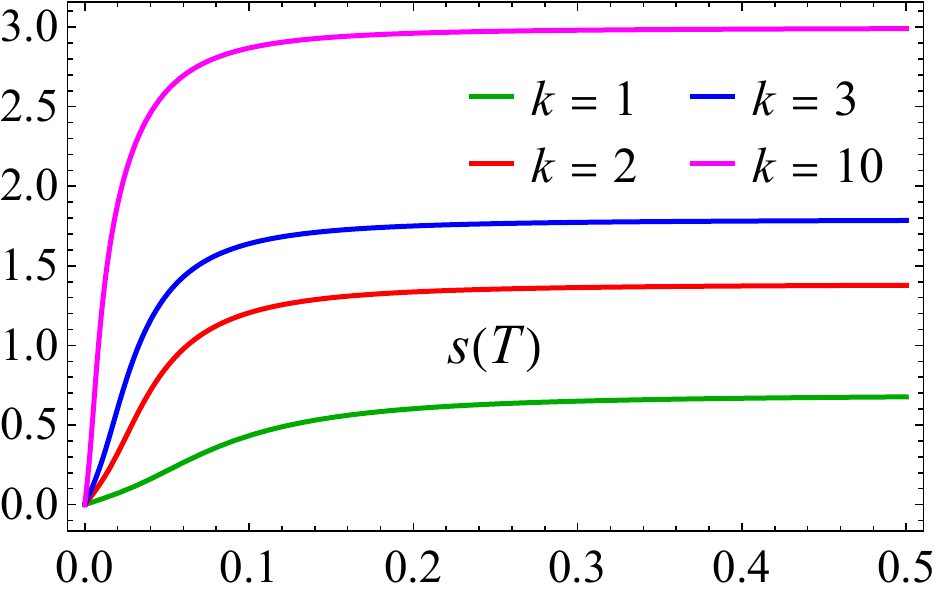}\hfill
  \includegraphics[height=.3\columnwidth]{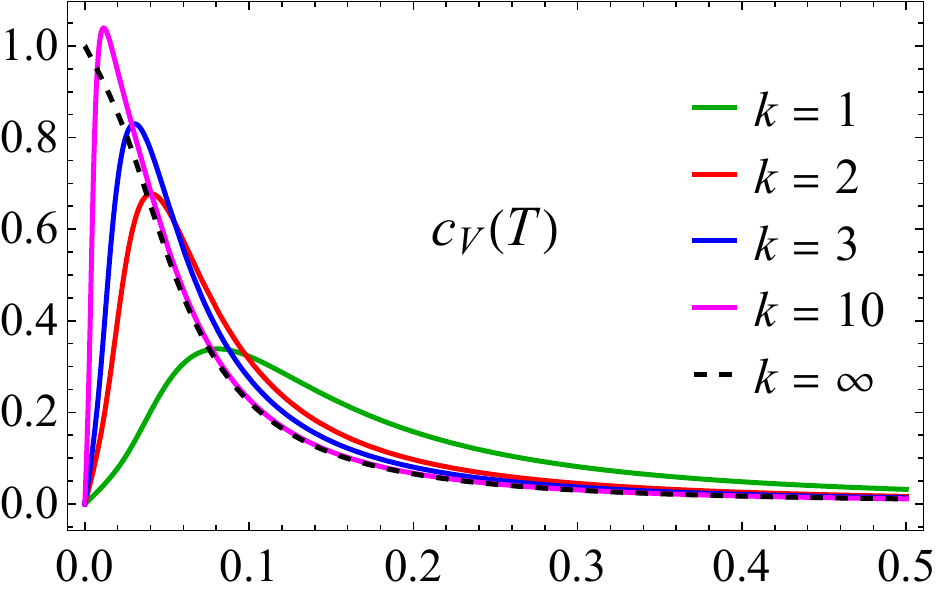}
  \caption{Thermodynamic free energy, energy, entropy, and specific heat per spin of the
    $\su(k|k)$ HS chain with $k=1,2,3,10$ (solid lines) vs.~temperature $T$ (in units of $J$). The
    dashed black line in the plots of $u$ and $c_V$ represents the asymptotic approximation to
    these functions given by Eqs.~\eqref{uapp}-\eqref{cVapp}.}
  \label{fig.thermokk}
\end{figure}%
\begin{remark}
  The integrals appearing in Eq.~\eqref{f11} for the $\su(k|k)$ free energy per spin can be
  readily evaluated in terms of the dilogarithm function (see Appendix~\ref{app.dilog}) for the PF
  chain, with the result
  \begin{multline}\label{fkkPF}
    f^{(k|k)}_{\text{PF}}(T)=-T^2\bigg[\vphantom{\frac{\pi^2}{12}}
    \Li_2(e^{-\be})+k\Big(\Li_2(-e^{-\be/k})\\
    -\Li_2(e^{-\be/k})\Big)+(3k-2)\frac{\pi^2}{12}\bigg].
  \end{multline}
  Differentiating this formula we get similar closed-form expressions for the energy, entropy and
  specific heat per spin. Note that, since $\Li_2(0)=0$, from the previous explicit formula for
  $f^{(k|k)}_{\text{PF}}$ we obtain
  \[
    f^{(k|k)}_{\text{PF}}(T)=-(3k-2)\frac{\pi^2T^2}{12}+o(T^2),
  \]
  which is consistent with the general asymptotic expression for the free energy per spin at low
  temperatures we shall derive in the next section.\qed
\end{remark}
As an example, we present in Fig.~\ref{fig.thermokk} a plot of the main thermodynamic functions of
the $\su(k|k)$ HS chain for several values of $k$ (the corresponding plots for the PF, FI and HS-B
chains are completely analogous and shall therefore be omitted). We have also compared in this
figure the energy and specific heat per unit volume computed using the exact formulas presented
above with their asymptotic approximations obtained by differentiating the asymptotic free energy
in Eq.~\eqref{fapp}, namely
\begin{align}
  u_\infty(T)&=T-\int_0^1\diff x\,\frac{\cE(x)}{e^{\be\cE(x)}-1},
               \label{uapp}\\
  c_{V,\infty}(T)&=1-\frac{\be^2}4\int_0^1\diff x\,\cE(x)^2\csch^2\bigl(\tfrac{\be\cE(x)}2\bigr).
          \label{cVapp}
\end{align}
As is apparent from Fig.~\ref{fig.thermokk}, this approximation is excellent even for $k=10$. We
have also analyzed the thermodynamic functions of the $\su(kp|kq)$ HS-type chains with
$(p,q)=(1,3), (3,1),(2,3),(3,2)$ using the explicit formulas for the Perron eigenvalue in these
cases listed above and Eq.~\eqref{f}, finding exactly the same qualitative behavior as in the case
$p=q=1$ discussed above. In fact, it should be noted that $u_\infty$ and $c_{V,\infty}$ do not
depend on $m$ and $n$, and thus provide an asymptotic approximation to $u^{(m|n)}$ and
$c_V^{(m|n)}$ for large $m$ and $n$. This implies that $u^{(m|n)}$ and $c_V^{(m|n)}$ become
virtually independent of $m$ and $n$ when both of these numbers are large.

\begin{figure}[h]
  \includegraphics[height=.3\columnwidth]{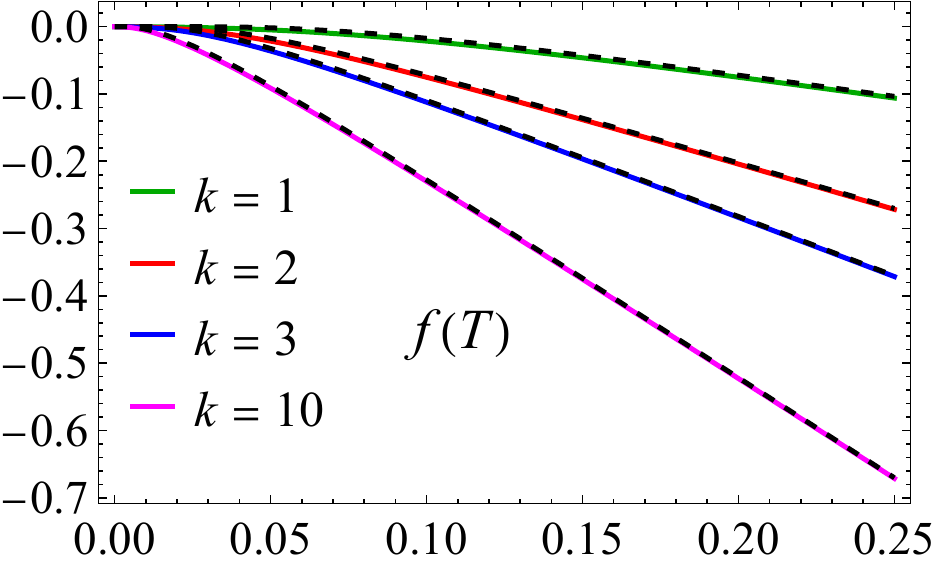}\hfill
  \includegraphics[height=.3\columnwidth]{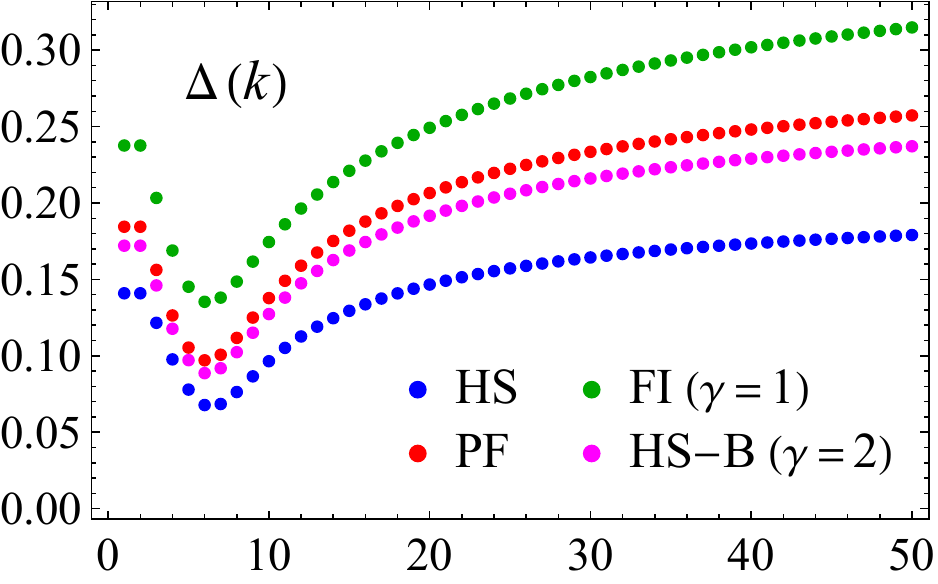}\\[3pt]
  \includegraphics[height=.3\columnwidth]{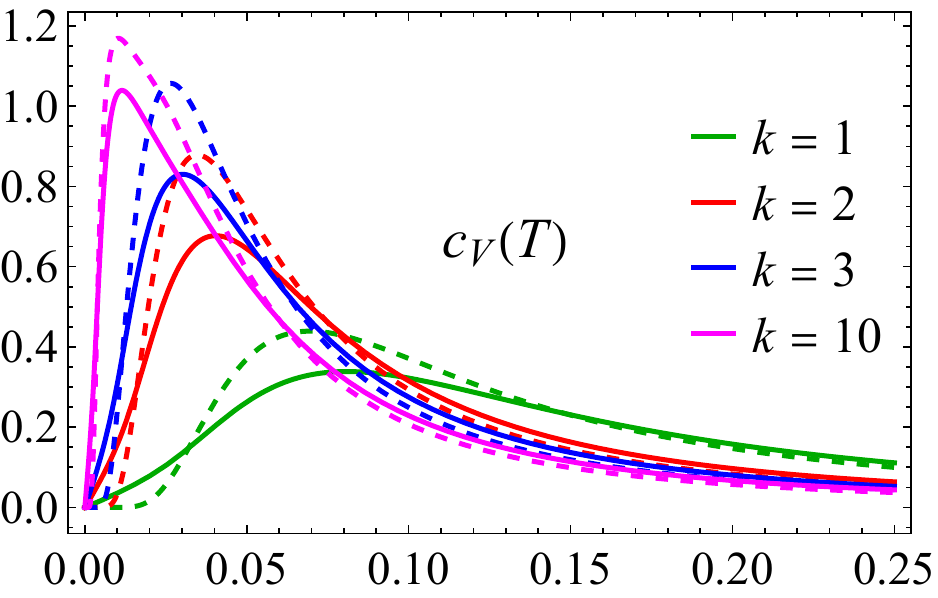}\hfill \null\en\kern3pt
  \includegraphics[height=.3\columnwidth]{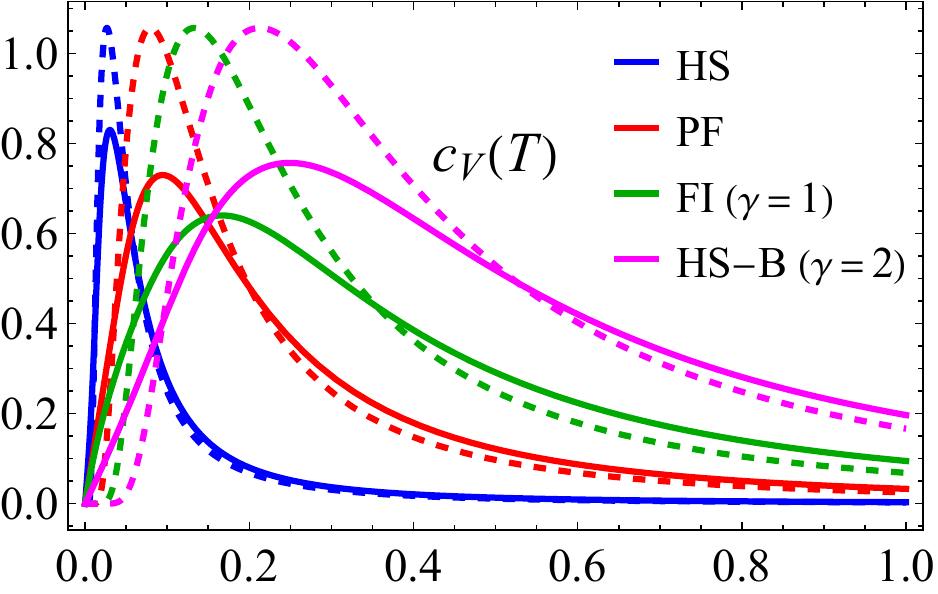}
  \caption{Top left: free energy per spin of the $\su(k|k)$ HS chain for several values of $k$
    (solid lines), compared to its $(k+1)$-level approximation~\eqref{fk} (dashed black lines).
    Top right: relative error $\De(k)$ of the latter approximation for each of the four families
    of $\su(k|k)$ HS-type chains with $k=1,\dots,50$. Bottom left: specific heat per spin of the
    $\su(k|k)$ HS chain with $k=1,2,3,10$ (solid lines) vs its $(k+1)$-level
    approximation~\eqref{cVk} (dashed lines). Top right: same comparison for the four families of
    $\su(k|k)$ HS-type chains with $k=3$. In all plots, the temperature is measured in units of
    $J$.}
  \label{fig.klevel}
\end{figure}

It is also apparent from Fig.~\ref{fig.thermokk} that the specific heat per spin of the $\su(k|k)$
HS chain features a single Schottky peak, i.e., a single pronounced global maximum, at
sufficiently low temperatures. The same is true for the other families of HS-type chains, as well
as for other values of $p$ and $q$ (see, e.g., the bottom panels in Fig.~\ref{fig.klevel}). For
the $\su(k|k)$ chains, the existence of a single Schottky peak can be qualitatively explained by
noting that when $T\gg\max\limits_{0\le x\le 1}\cE(x)$ we can approximate $e^{-\be\cE(x)}$ in
Eq.~\eqref{f} for $f(T)$ by $e^{-\be\cE_0}$, which amounts to approximating $\cE(x)$ by its
average $\cE_0$ over the interval $[0,1]$. In this way we obtain the approximation
\[
  f(T)\sim-T\ln\left(\la_1[e^{-\be\cE_0}]\right),
\]
which for the $\su(k|k)$ chain reduces to
\begin{align}
  f(T)&\sim-T\ln\left(\frac{1-e^{-\be\cE_0}}{1-e^{-\be\cE_0/k}}\,(1+e^{\be\cE_0/k})\right)\notag\\
      &=\ln\biggl(1+2\sum_{j=1}^{k-1}e^{-\frac{j}{k}\be\cE_0}+e^{-\be\cE_0}\biggr)\equiv f_k(T),
        \label{fk}
\end{align}
which is the free energy of a $(k+1)$-level system with energies $E_j=j\cE_0/k$ (with
$j=0,\dots,k$) and degeneracies
\begin{equation}\label{klevdeg}
  g_0=g_k=1,\qquad g_j=1\en (1\le j\le k-1).
\end{equation}
The agreement of the $(k+1)$-level approximation~\eqref{fk} with the exact value of the free
energy per spin is quite good even at low temperature (cf.~Fig.~\ref{fig.klevel}, top). The
specific heat per spin of the $(k+1)$-level system, given by
\begin{multline}
  \label{cVk}
  c_{V,k}(T)=\be^2\cE_0^2\left(\frac1{k^2}\csch\bigl(\tfrac{\be\cE_0}{k}\bigr)\right.
    \coth\bigl(\tfrac{\be\cE_0}{k}\bigr)\\
    \left.-\frac14\csch^2\bigl(\tfrac{\be\cE_0}{2}\bigr)\right),
\end{multline}
features a single Schottky peak whose temperature provides a reasonable approximation to the
temperature of the Schottky peak of the corresponding $\su(k|k)$ chain (see, e.g., the bottom of
Fig.~\ref{fig.klevel}).

\section{Low temperature behavior and criticality}\label{sec.CB}

In this section we shall analyze the critical behavior of the $\su(m|n)$ supersymmetric HS type
chains~\eqref{HANm1}-\eqref{Bchain} in the genuinely supersymmetric case $mn\ne0$. As mentioned in
the Introduction, a necessary condition for criticality is the low-temperature behavior~\eqref{fc}
of the thermodynamic free energy per spin, characteristic of a $(1+1)$-dimensional CFT with
central charge $c$. We shall therefore start by showing that Eq.~\eqref{fc} indeed holds in all
cases (the only exception being the FI chain with $\ga=0$), with $f(0)=0$ (cf.~the first remark at
the end of Section~\ref{sec.fe}) and
\begin{equation}
  \label{cc}
  c=m+\frac{n}2-1.
\end{equation}

The proof of Eq.~\eqref{fc} relies on the general formula~\eqref{f} expressing the thermodynamic
free energy per spin in terms of the Perron eigenvalue $\la_1$ of the $(m+n)\times(m+n)$ transfer
matrix $A(x)$ with matrix elements~\eqref{Amat}. The key idea is to perform the change of variable
\begin{equation}\label{yvar}
  y=J\be\cE(x)
\end{equation}
in Eq.~\eqref{f}. Note that the dispersion function $\cE(x)$ is monotonically increasing over the
whole integration interval $[0,1]$ for the PF, FI, and HS-B chains. However, for the HS chain
$\cE(x)=x(1-x)$ is one-to-one only in the interval $[0,1/2]$, and is symmetric about the midpoint
$x=1/2$. For this reason, we rewrite Eq.~\eqref{f} as
\[
  f(T)=-\frac{T}{\eta}\int_0^\eta\diff x\,\ln\bigl(\la_1[a]\bigr),
\]
where $\eta=1/2$ for the HS chain and $\eta=1$ for the remaining chains of HS type. Note that in
the previous formula we have also stressed the fact that the Perron eigenvalue $\la_1$ depends on
the variable $x$ through $a\equiv e^{-J\be\cE(x)}$. Performing the change of variable~\eqref{yvar}
in the latter equation we then obtain
\[
  f(T)=-\frac{T^2}{\eta J}\int_0^{J\be\cE(\eta)}\frac{\diff y}{\cE'(x)}\ln\left(\la_1[e^{-y}]\right).
\]
It is shown in Appendix~\ref{app.int} that the leading behavior as $T\to\infty$ of the integral in
the right-hand side of the previous equation is obtained replacing~\footnote{We must exclude the
  value $\ga=0$ of the parameter in the FI chain to guarantee that $\cE'(0)\ne0$.} $\cE'(x)$ by
$\cE'(0)$ and pushing the upper integration limit to $\infty$. In other words,
\begin{equation}\label{fcapp}
  f(T)=-\frac{T^2}{\eta J\cE'(0)}\int_0^{\infty}\diff y\ln\left(\la_1[e^{-y}]\right)+o(T^2).
\end{equation}
Since the integral in the last equation is independent of $T$, this shows that the free energy per
spin of the supersymmetric HS-type chains~\eqref{HANm1}-\eqref{Bchain} behaves as
$-\text{const.}T^2$ as $T\to0+$. To compute the constant $c$ in Eq.~\eqref{fc}, we must however
determine the Fermi velocity $v_F$. Although this will be done rigorously later in this section by
studying the low energy excitations above the ground state, we can determine $v_F$ heuristically
by noting that the integration range $x\in[0,\eta]$ should correspond to the interval $[0,\pi]$ in
momentum space. In other words, we should have
\[
  x=\frac{\eta p}\pi,
\]
and therefore
\begin{equation}\label{vF}
  v_F=\frac{\diff}{\diff p}\bigg|_{p=0}J\cE(x)=\frac{\eta}\pi J\cE'(0).
\end{equation}
We can therefore rewrite Eq.~\eqref{fcapp} as
\[
  f(T)=-\frac{T^2}{\pi v_f}\int_0^{\infty}\diff y\ln\left(\la_1[e^{-y}]\right)+o(T^2),
\]
which yields the following exact formula for the central charge $c$:
\begin{equation}
  \label{ccgen}
  c=\frac6{\pi^2}\int_0^{\infty}\diff y\ln\left(\la_1[e^{-y}]\right).
\end{equation}
Remarkably, although the Perron eigenvalue $\la_1$ can only be computed in closed form in a
handful of cases, it is possible to evaluate the last integral using a variant of the technique
developed in Ref.~\cite{HB00} to study the PF chain. Indeed, let us perform in the last equation
the change of variables $\al=e^{-y}$, so that
\[
  c=\frac6{\pi^2}\int_0^1\frac{\diff\al}{\al}\ln\left(\la_1[\al]\right).
\]
Setting
\[
  \mu(\al)=1-\frac{1-\al}{\la_1[\al]}
\]
and using Eq.~\eqref{Li21}, we can rewrite the previous equation as
\begin{multline}
  c=\frac6{\pi^2}\int_0^1\frac{\diff\al}{\al}\ln\left(\frac{1-\al}{1-\mu(\al)}\right)\\
  =-1-\frac6{\pi^2}\int_0^1\frac{\diff\al}{\al}\ln\bigl(1-\mu(\al)\bigr).
  \label{clast}
\end{multline}
At this point we make use of the fact that the Perron eigenvalue satisfies the eigenvalue
Eq.~\eqref{eigveq} (with $(p,q)$ replaced by $(m,n)$), which in terms of $\mu$ can be recast as
\begin{equation}\label{almu}
  \al=\frac{\mu(\al)^m}{\big(2-\mu(\al)\big)^n}.
\end{equation}
Since
\[
  \frac1\al\frac{\diff\al}{\diff\mu}=\frac{m}{\mu}+\frac{n}{2-\mu}>0
\]
(where we have used the fact that $\la_1>1$ implies $\mu\in[0,1]$), Eq.~\eqref{almu} is a valid
change of variables. Performing this change of variable in the integral in Eq.~\eqref{clast} we
obtain:
\begin{align*}
  -\int_0^1\frac{\diff\al}{\al}
  &\ln\bigl(1-\mu(\al)\bigr)\\
  &=
    -\int_0^1\diff\mu\left(m\,\frac{\ln(1-\mu)}\mu+n\,\frac{\ln(1-\mu)}{2-\mu}\right)\\
  &=\frac{\pi^2}6\left(m+\frac{n}2\right)
\end{align*}
(cf.~Appendix~\ref{app.dilog} for the evaluation of the last integral), which substituted into
Eq.~\eqref{clast} yields Eq.~\eqref{cc}.
\begin{remark}
  In the case of the FI chain with $\ga=0$, proceeding as before we obtain the asymptotic estimate
  \[
    f(T)=-\ka T^{3/2}+o(T^{3/2}),
  \]
  with
  \[
    \ka=\frac1{2\sqrt{J}}\int_0^\infty\frac{\diff y}{\sqrt y}\,\ln\bigl(\la_1[e^{-y}]\bigr).
  \]
  Hence this chain is not critical.\qed
\end{remark}
Although the low temperature behavior~\eqref{fc} of the free energy per spin is a necessary
condition for criticality, it is not sufficient. We must still require that the spectrum have a
finite degeneracy as the number of spins $N$ goes to infinity, and that in this limit there exist
gapless low energy excitations above the ground state with a linear energy-momentum relation. For
the $\su(m|n)$ HS chain, this was shown to be the case only for $m=0$ (purely fermionic, or
antiferromagnetic case) and $m=1$~\cite{BBS08}. In Ref.~\cite{FG22pre} this result was extended to
all four families of HS-type chains in the non-supersymmetric case $mn=0$. We shall therefore
restrict ourselves in what follows to the PF, FI and HS-B chains in the truly supersymmetric case
$mn\ne0$.

Consider first the PF and FI chains, for which the sets $B$ and $F$ of bosonic and fermionic
degrees of freedom are given by Eq.~\eqref{BFA}. Since the dispersion relation~\eqref{disp} of the
latter chains is monotonically increasing, the ground state of the spectrum~\eqref{bondE} is
obtained from bond vectors $\bsv=(s_1,\dots,s_{N})\in\{1,\dots,m+n\}$ whose corresponding motifs
$(\de(s_1,s_2),\dots,\de(s_{N-1},s_N))$ consist of the highest possible number of $0$'s starting
from the right end. Therefore the ground state is the zero mode, obtained from bond vectors
$\bsv=(s_1,\dots,s_N)$ of the form
\[
  s_1\le\cdots\le s_{N-k}<s_{N-k+1}<\cdots<s_N
\]
with $s_i\in B$ for $1\le N-k$, $s_i\in F$ for $N-k+1\le i\le N$ and $k=0,\dots,n$. The ground
state degeneracy is thus
\begin{equation*}
  g=\sum_{k=0}^n\binom{N-k+m-1}{m-1}\binom{n}{k}.
\end{equation*}
For $m>1$ we have
\[
  g\ge \binom{N+m-1}{m-1}\ge\frac{N^{m-1}}{(m-1)!},
\]
which diverges when $N\to\infty$. It follows that the $\su(m|n)$ PF and FI chains with $m>1$
cannot be critical. On the other hand, for $m=1$ the ground state degeneracy $g=2^n$ remains
finite (constant) as $N\to\infty$.

A similar analysis can be performed for the HS-B chain, taking into account that in this case the
sets $B$ and $F$ are given by Eq.~\eqref{BFB}, and the spectrum is obtained from bond vectors
$\bsv=(s_1,\dots s_N,s^*)$ (with $s^*$ defined by Eq.~\eqref{sNp1}) through Eq.~\eqref{EbsvB}.
Thus for $m_{\vep_B}\ge1$ (i.e., $m\ge2$ or $m=1$ and $\vep_B=1$) proceeding as before it can be
readily shown that the ground state degeneracy is given by
\[
  g=\sum_{k=0}^{n_{\vep_F}}\binom{N-k+m_{\vep_B}-1}{m_{\vep_B}-1}\binom{n_{\vep_F}}{k}\ge
  \binom{N+m_{\vep_B}-1}{m_{\vep_B}-1},
\]
which diverges as $N\to\infty$ for $m_{\vep_B}\ge2$. It follows that the $\su(m|n)$ HS-B chain
with $m_{\vep_B}\ge2$, i.e., $m\ge 4$ or $m=3$ and $\vep_B=1$, cannot be critical. On the other
hand, for $m_\vep=1$, i.e., $(m=1,\vep_B=1)$, $m=2$, and $(m=3,\vep_B=-1)$, the ground state
degeneracy is $2^{n_{\vep_F}}$, and thus remains finite as $N\to\infty$. Likewise, when $m=1$ and
$\vep_B=-1$ we have $m_{\vep_B}=0$, and therefore $F=\{1,\dots,n\}$ and $B=\{n+1\}$ according to
Eq.~\eqref{BFB}. The ground state, which is obtained from bond vectors of the form
\[
  (s_1,\dots,s_k,\,\underbrace{n+1,\dots,n+1}_{N-n_{\vep_F}-k}\,,1,2,\dots,n_{\vep_F})
\]
with $1\le s_1<\cdots <s_k\le n$ and $k=0,\dots,n$, has
energy~$\cE(1-n_{\vep_F}/N)=\ga+\frac12+O(N^{-1})$ and finite (constant) degeneracy $2^n$.

From the previous analysis of the ground state degeneracy it follows that the only $\su(m|n)$
(truly) supersymmetric chains of HS-type that can be critical are the HS, PF and FI chains with
$m=1$ (and $\ga>0$ for the FI chain), and the HS-B chain with $m=1$, $m=2$, and $(m=3,\vep_B=-1)$.
As mentioned above, the $\su(1|n)$ HS chain was shown to be critical in Ref.~\cite{BBS08}. The
$\su(1|n)$ PF and FI chain (with $\ga>0$ in the latter case) can be dealt with in essentially the
same fashion, as we shall next prove. Indeed, in this case the ground state (zero mode) bond
vectors are of the form
\[
  \bigl(\,\underbrace{1,\dots,1}_{N-k}\,,s_{N-k+1},\dots,s_N\bigr)
\]
with $2\le s_{N-k+1}<\cdots< s_N\le n$ and $k=0,\dots,n$. Gapless low energy excitations above the
ground state are obtained, for instance, by changing the $1$ in the $l$-th component of the
previous bond vector into (for instance) $m+n$, provided that $l\ll N$. Indeed, the energy of the
resulting bond vector
\begin{equation}\label{bsv}
  \bsv=(1,\dots, 1,\underset{\underset l\uparrow}{m+n},1,\dots,1,s_{N-k+1},\dots,s_N)
\end{equation}
is
\[
  \De E=J\cE(l/N)\simeq\frac{Jl}N\cE'(0)=O(N^{-1}),
\]
since $l\ll N$ and $\cE'(0)\ne0$. Although the PF and FI chains are not translationally invariant,
and thus linear momentum is not conserved, for a state with bond vector $\bsv$ we can define a
pseudo-momentum $P$ by
\[
  P(\bsv)=\sum_{i=1}^N\frac{\pi i}N\,\de(s_i,s_{i+1})\quad \mod 2\pi.
\]
Note that the factor of $\pi$ (instead of the usual one of $2\pi$, cf.~\cite{HHTBP92,BBH10}) is
due to the fact that the dispersion relation $\cE(x)$ of the PF and FI chains is monotonically
increasing over the interval $[0,1]$, and thus this interval represents only the positive momentum
sector $0\le p\le\pi$. With this definition the ground states have zero momentum, and the momentum
of the low energy excitation with bond vector~\eqref{bsv} is
\[
  \De P=\frac{\pi l}N.
\]
Thus the energy-momentum relation of the latter excitation is linear, with Fermi velocity
\[
  v_F=\frac{\De E}{\De p}=\frac{J\cE'(0)}{\pi},
\]
in agreement with Eq.~\eqref{vF} (recall that $\eta=1$ for the PF, FI and HS-B chains). It follows
that the $\su(1|n)$ PF and FI chains are critical. A totally similar argument shows that the same
is true for the $\su(m|n)$ HS-B chain with $m=1,2,3$ (provided that $\vep_B=-1$ in the latter
case).

\section{Conclusions}\label{sec.conc}

In this paper we have studied the thermodynamics of four families of $\su(m|n)$ supersymmetric
spin chains of Haldane--Shastry type in the genuinely supersymmetric case $mn\ne0$. It was
previously known that the thermodynamic free energy per spin of all of these models can be
expressed in a unified way in terms of the Perron (i.e., largest in modulus) eigenvalue of an
appropriate inhomogeneous transfer matrix $A(x)$, where $x\in[0,1]$ is proportional to the
effective linear momentum of the excitation considered (see, e.g.,~\cite{FGLR18,FG22jstat}). In a
previous publication~\cite{FG22pre}, we used this idea to evaluate in closed form the free energy
per spin and hence the main thermodynamic functions of the $\su(m|0)$ and $\su(0|m)$ (i.e.,
non-supersymmetric) chains for arbitrary $m=2,3,\dots$. In the non-supersymmetric case, however,
due to the lack of symmetry of the transfer matrix there were so far exact results only for
$m,n\le 2$. In this work we have derived a remarkable relation expressing the Perron eigenvalue of
an $\su(kp|kq)$ chain with arbitrary $k=1,2,\dots$ in terms of the Perron eigenvalue of the
$\su(p|q)$ chain. Applying this relation to the cases mentioned above for which the $\su(p|q)$
free energy per spin had been explicitly computed, we have been able to determine in closed form
the free energy per spin of the $\su(k|k)$, $\su(2k|k)$, and $\su(k|2k)$ chains for arbitrary
$k=1,2,\dots$. We have also evaluated the Perron eigenvalue in the $\su(1|3)$, $\su(3|1)$,
$\su(2|3)$ and $\su(3|2)$ cases, and in this way computed in closed form the free energy per spin
of the corresponding infinite families of $\su(k|3k)$, $\su(3k|k)$, $\su(2k|3k)$ and $\su(3k|2k)$
chains with $k=1,2,\dots$.

The results mentioned above make it possible to study the thermodynamics of the infinite families
of $\su(m|n)$ supersymmetric spin chains of HS type whose free energy per spin we have computed in
closed form. Our analysis shows that the thermodynamic behavior of these models is qualitatively
similar to that of their non-supersymmetric counterparts. In particular, the specific heat per
unit volume features a single marked Schottky peak at relatively low temperatures. In the case of
the $\su(k|k)$ chains, for which the exact formulas for the main thermodynamic formulas are
especially simple, we have been able to heuristically justify this behavior by showing that the
free energy per spin of these chains can be roughly approximated by the free energy of a simple
$(k+1)$-level model. By studying the specific heat of the latter model we have been able to
justify the existence of a single Schottky peak in the specific heat of $\su(k|k)$ chains and
obtain a rough approximation to its temperature.

Another novel result which follows from our study of the Perron eigenvalue of the $\su(m|n)$
transfer matrix is the fact that the energy and specific heat per unit spin of the models under
study tend to well defined limits as $m$ and $n$ tend to infinity, which are moreover independent
of $m$ and $n$. In fact, our numerical calculations indicate that this asymptotic regime is
achieved for values of $m$ and $n$ as low as $10$.

A key property of $\su(m|n)$ supersymmetric spin chains of HS type that we have analyzed in this
work is their critical behavior. As mentioned in the Introduction, the deep connections between
the latter models and conformal field theories in $(1+1)$ dimensions were detected very early on,
starting with the seminal work of Haldane and collaborators relating the spectrum of the original
HS chain to the $\su(2)_1$ WZNW model~\cite{HHTBP92}. As is well known, the free energy of a
$(1+1)$ dimensional CFT, and thus of any critical finite model, must be asymptotically
proportional to $-T^2$ at low temperatures. Determining the proportionality factor is in fact of
great interest, since it yields the central charge $c$ of the theory. We have studied the low
temperature behavior of the free energy per spin of all four families of $\su(m|n)$ supersymmetric
spin chains of HS type, showing that in all cases (with the only exception of the FI chain with
$\ga=0$) it follows the asymptotic law~\eqref{fc} with central charge $c=m-1+n/2$. However, this
low-temperature asymptotic behavior of the free energy is not enough to guarantee that a finite
model is critical. Indeed, the ground state of a critical model must have finite degeneracy in the
thermodynamic limit, and there must be gapless low energy excitations with a linear
energy-momentum dispersion relation. Using the description of the spectrum of the chains under
study in terms of (generalizations of) supersymmetric Haldane motifs, we have been able to
complete earlier work on the supersymmetric HS chain~\cite{BBS08}, and classify all (genuinely)
supersymmetric chains of HS type that are truly critical. More precisely, we have found that the
PF and FI chains are only critical when $m=1$, as is the case with the HS chain, while the HS-B
chain is also critical when $m=2$ and $m=3$ (provided that in the latter case the sign $\vep_B$ in
the Hamiltonian takes the value $-1$).

The present work leaves several open problems that may be the subject of future research. In the
first place, although our results reduce the computation of the thermodynamic free energy per spin
of $\su(m|n)$ supersymmetric spin chains of HS type to the case of $m$ and $n$ relatively prime,
the latter case remains still intractable except for very low values of $m$ and $n$. In fact, the
exact expressions of the Perron eigenvalue we have obtained in this work for $m,n\le 3$ do not
suggest any simple ansatz for higher values of $m$ and $n$ due to their complexity. On the other
hand, the novel closed-form expression for the characteristic polynomial of the $\su(m|n)$
transfer matrix for arbitrary values of $m$ and $n$ derived in this work is the natural starting
point for computing the Perron eigenvalue, and thus the main thermodynamic functions, for larger
values of $m$ and $n$. Another application of our results that could be worth exploring is the
determination of the thermodynamical properties of the new class of solvable translation-invariant
$\su(m|n)$ supersymmetric spin chains with long-range interactions introduced in
Ref.~\cite{FG22jstat}. Indeed, when $m$ and $n$ are both even the free energy of these models was
shown to be the sum of the free energies of the $\su(1|1)$ and
$\su\bigl(\frac{m}2|\frac{n}2\bigr)$ HS chains, and thus can be exactly determined in a few cases
using the exact formulas in Section~\ref{sec.fe}.

\section*{Acknowledgments}
This work was partially supported by grant~GRFN24/24 from Universidad Complutense de Madrid.

\appendix
\section{Supersymmetric permutation and spin reversal operators}\label{app.Sij}

In this appendix we shall briefly review the definition of the supersymmetric spin permutation and
reversal operators. For the sake of convenience, we shall respectively denote by
$B=\{b_1,\dots,b_m\}$ and
\[
  F=\{1,\dots,m+n\}\setminus B=\{f_1,\dots,f_n\}
\]
the sets of bosonic and fermionic degrees of freedom. The system's Hilbert space is the set
$\cH=(\CC^{m+n})^{\otimes N}$ spanned by the canonical basis vectors
\[
  \ket{s_1\cdots s_N}\equiv\ket{\bbs}:=a^\dagger_{1s_1}\cdots a^\dagger_{Ns_N}\ket{0},\quad
  s_i\in B\cup F,
\]
where $\ket0$ is the Fock vacuum and the operator $a^\dagger_{i\si}$ acts on the $i$-th site and
creates a boson (respectively a fermion) of type $\si$ for $\si\in B$ (respectively $\si\in F$).
We thus have
\begin{align*}
  \big[a_{i\al},a_{j\be}\big]
  &=0,\quad
    \big[a_{i\al},a_{j\be}^\dagger\big]=\de_{ij}\de_{\al\be},
    \qquad \al,\be\in B,\\
  \big\{a_{i\al},a_{j\be}\big\}
  &=0,\quad \big\{a_{i\al},a_{j\be}^\dagger\big\}=\de_{ij}\de_{\al\be},\qquad \al,\be\in F,\\
  \big[a_{i\al},a_{j\be}\big]
  &=\big[a_{i\al},a_{j\be}^\dagger\big]=
    0,\qquad i\ne j,\enspace \al\in B,\enspace \be\in F.
\end{align*}
The supersymmetric permutation operator $S_{ij}$ is naturally defined as
\[
  S_{ij}=\sum_{\al,\be=1}^{m+n}a_{i\be}^\dagger a_{j\al}^\dagger a_{i\al}a_{j\be}.
\]
It immediately follows from this definition that in the purely bosonic case $n=0$ the
operator $S_{ij}$ reduces to the ordinary spin permutation operator $P_{ij}$ defined by
\[
  P_{ij}\ket{\cdots s_i\cdots s_j\cdots}=\ket{\cdots s_j\cdots s_i\cdots},
\]
while in the purely fermionic case $m=0$ we have $S_{ij}=-P_{ij}$. Let us define the grading
$p:B\cup F\to\{0,1\}$ as the characteristic function of the fermionic set $F$, i.e., $p(\si)=0$
for $\si\in B$ and $p(\si)=1$ for $\si\in F$. In general (assuming, without loss of generality,
that $i<j$), we can write $S_{ij}=\vep_{ij}P_{ij}=P_{ij}\vep_{ij}$, with
$\vep_{ij}\ket{\bbs}=p(s_i)\ket\bbs$ when $p(s_i)=p(s_j)$ and
$\vep_{ij}\ket{\bbs}=(-1)^{\nu_{ij}}\ket\bbs$ if $p(s_i)\ne p(s_j)$, where
\[
  \nu_{ij}=\sum_{k=i+1}^{j-1}p(s_k).
\]
It is also straightforward to check that the supersymmetric spin permutation operators $S_{ij}$
obey the standard permutation algebra relations
\[
  S_{ij}^2=1,\quad S_{ij}S_{jk}=S_{jk}S_{ik}=S_{ik}S_{ij},\quad S_{ij}S_{kl}=S_{kl}S_{ij},
\]
where $i,j,k,l$ are distinct indices.

We shall next define the supersymmetric spin reversal operators $S_i$. To this end, we introduce
the spin flip mapping~$\si\mapsto\si'$ given by
\[
  b'_{\al}=b_{m+1-\al},\qquad f'_{\al}=f_{n+1-\al}.
\]
In other words, ${'}$ maps the bosonic spin degrees of freedom $b_1,\dots,b_m$ into
$b_m,\dots,b_1$, and similarly for the fermionic ones. We then define
\[
  S_i=\sum_{\si=1}^{m+n}\vep(\si)a^\dagger_{i\si'}a_{i\si},
\]
where $\vep(\si)=\vep_B\in\{\pm1\}$ for $\si\in B$ and $\vep(\si)=\vep_F\in\{\pm1\}$ for
$\si\in F$. In particular, we have $S_i^2=1$ regardless of the choice of the signs $\vep_{B,F}$.

\section{Eigenvalues of the transfer matrix}\label{app.Adiag}

In this appendix we shall prove a key property of the eigenvalues of the transfer matrix used in
Section~\ref{sec.fe} to relate the free energies of the $\su^{(p|q)}$ and $\su^{(kp|kq)}$ chains
with $k=1,2,\dots\,.$ More generally, consider a matrix
$M=(m_{ij})_{1\le i,j\le n}\equiv M(\al,\be,\bga)$ of the form
\[
  m_{ij}=
  \begin{cases}
    \al,& i<j\\
    \ga_i,& i=j\\
    \be,& i>j.
  \end{cases}
\]
We shall show that
\begin{equation}\label{dets}
  \det M(\al,\be,\bga)=\det M(\al,\be,\tilde\ga),
\end{equation}
where $\tilde\ga$ is obtained from $\ga$ by permuting its components. Applying this result to the
matrix $M(\al,\be,\bga)-\la\id$, which has the same structure as $M(\al,\be,\bga)$, implies that
the eigenvalues of the latter matrix are invariant under permutations of its diagonal elements.

To begin with, note that it suffices to prove Eq.~\eqref{dets} when $\tilde\ga$ is differs from
$\ga$ by permuting two consecutive components, for instance the first and the second one. The only
terms in $\det M(\al,\be,\bga)$ not invariant under exchange of $\ga_1$ and $\ga_2$ are
\begin{multline}\label{ga1ga2}
  \sum_\si(-1)^\si\ga_{1}m_{2\si_2}\cdots
  m_{n\si_n}\\+\sum_\rho(-1)^\rho\ga_{2}m_{1\rho_1}m_{3\rho_3}\cdots m_{n\rho_n},
\end{multline}
where $\si=(1,\si_2,\dots,\si_n)$ and $\rho=(\rho_1,2,\rho_3,\dots,\rho_n)$ are permutations of
$(1,\dots,n)$ with $\si_2,\rho_1\ge 3$, and $(-1)^\pi$ denotes the sign of the permutation $\pi$.
To prove the invariance of $\det M(\al,\be,\bga)$ under exchange of $\ga_1$ and $\ga_2$, we shall
check that these two terms are mapped into one another under such an exchange. Indeed, note first
of all that $m_{2\si_2}=m_{1\si_2}=\al$ as $\si_2\ge3$. Likewise, if $\si_j=2$ for some
$j\in\{3,\dots,n\}$ then $m_{j\si_j}=m_{j2}=m_{j1}=\be$. Hence
\begin{align*}
  (-1)^\si&\ga_{1}m_{2\si_2}\cdots m_{n\si_n}\\
          &=(-1)^\si\ga_{1}m_{1\si_2}m_{3\si_3}\cdots
            m_{j1}\cdots m_{n\si_n}\\
          &=(-1)^\rho\ga_{1}m_{1\rho_1}m_{3\rho_3}\cdots
            m_{j1}\cdots m_{n\si_n},
\end{align*}
with
\[
  \rho=(\si_2,2,\si_3,\dots,1,\dots,\si_n)
\]
a permutation of $(1,\dots,n)$ with $\rho_1=\si_2\ge3$. This shows that the two terms in
Eq.~\eqref{ga1ga2} are mapped into one another under the exchange of $\ga_1$ and $\ga_2$, thus
establishing our claim.

\section{Estimate of an integral}\label{app.int}

We wish to prove the asymptotic estimate
\begin{multline}\label{asyrel}
  \int_0^{J\be\cE(\eta)}\frac{\diff y}{\cE'(x)}\ln\left(\la_1[e^{-y}]\right)\\
  =\frac1{\cE'(0)}\int_0^\infty\diff y\ln\left(\la_1[e^{-y}]\right)+o(1).
\end{multline}
which we used in Section~\ref{sec.CB} to prove Eq.~\eqref{ccgen} for the central charge. In the
case of the PF, FI and HS-B chains ---with $\ga>0$ for the latter two chains--- this is
straightforward, since we can write
\begin{align}
  &\int_0^{J\be\cE(1)}\frac{\diff y}{\cE'(x)}\ln\left(\la_1[e^{-y}]\right)
    -\frac1{\cE'(0)}\int_0^\infty\diff y\ln\left(\la_1[e^{-y}]\right)
    \notag\\
  &=\int_0^{J\be\cE(1)}\diff y\left(\frac1{\cE'(x)}-\frac1{\cE'(0)}\right)
    \ln\left(\la_1[e^{-y}]\right)\notag\\
  &\hphantom{\frac1{\cE'(0)}\int^{J\be\cE(1)}\diff y}
    -\frac1{\cE'(0)}\int_{J\be\cE(1)}^\infty\diff y\ln\left(\la_1[e^{-y}]\right).
    \label{estint}
\end{align}
The second term in the right-hand side of Eq.~\eqref{estint} is clearly $o(1)$ as $\be\to\infty$,
as the integral
\[
  \int_0^\infty\diff y\ln\left(\la_1[e^{-y}]\right)
\]
is convergent (its value was actually computed in Section~\ref{sec.CB}). To see that also the
first term is $o(1)$, note that
\[
  \frac1{\cE'(x)}-\frac1{\cE'(0)}\equiv\frac1{\cE'\bigl(\cE^{-1}(Ty)\bigr)}-\frac1{\cE'(0)}=O(Ty),
\]
since $\cE'$ does not vanish on the interval $[0,1]$ for the above mentioned chains and
$\cE^{-1}(0)=0$ (here $\cE^{-1}$ denotes the inverse function of the restriction of $\cE$ to the
interval $[0,1]$). We thus have
\begin{multline}\label{cEp}
\left|\int_0^{J\be\cE(1)}\diff y\left(\frac1{\cE'(x)}-\frac1{\cE'(0)}\right)
  \ln\left(\la_1[e^{-y}]\right)\right|\\
\le CT\int_0^{J\be\cE(1)}\diff y\,y\ln\left(\la_1[e^{-y}]\right)
\end{multline}
for some positive constant $C$. We shall show below that
\[
  \la_1[e^{-y}]=1+O(e^{-y/m})
\]
as $y\to\infty$ (see Eq.~\eqref{Puis}), which implies that the integral
$\int_0^\infty\diff y\,y\ln\left(\la_1[e^{-y}]\right)$ is convergent. Therefore the right-hand
side of Eq.~\eqref{cEp} is $O(T)$ (and, in particular, $o(1)$).

A finer argument is needed in the case of the HS chain and the HS-B chain with $\ga=0$, for which
$\cE'$ vanishes at the upper end $\eta$ of the integration interval. In fact, since for $\ga=0$
the free energy per spin of these two chains are related by
\begin{align*}
  f_{\mathrm{HS-B}}(T;J)
  &=-T\int_0^1\diff x\,\ln\bigl(\la_1[e^{-J\be x(1-\frac{x}2)}]\bigr)\\
  &=-2T\int_0^{1/2}\diff s\,\ln\bigl(\la_1[e^{-2J\be s(1-s)}]\bigr)\\
  &=f_{\mathrm{HS}}(T;2J),
\end{align*}
we can restrict our discussion to the HS chain. In this case we have
\[
  \cE'(x)=1-2x=\left(1-\frac{4y}{\be J}\right)^{1/2},
\]
where we have used Eqs.~\eqref{disp} and~\eqref{yvar} to express $x$ in terms of the independent
variable $y$. Since the last term in the right-hand side of Eq.~\eqref{estint} is $o(1)$, all we
have to prove in this case is that
\begin{equation}\label{asyint}
  \int_0^{J\be/4}\diff y\left[\left(1-\frac{4y}{\be J}\right)^{-1/2}-1\right]
  \ln\left(\la_1[e^{-y}]\right)
\end{equation}
tends to zero as $\be\to\infty$. To prove our claim, note that we can write the last integral as
\begin{multline*}
  \int_0^{J\be/4}\diff y\,h'(y/\be J)\ln\left(\la_1[e^{-y}]\right)\\
  =
  \be J\int_0^{J\be/4}\diff y\,\ln\left(\la_1[e^{-y}]\right)\frac\diff{\diff y}\,h(y/\be J),
\end{multline*}
with
\[
  h(z)=\frac12\left(1-2z-\sqrt{1-4z}\,\right)
\]
and $h'(z)=\frac{\diff h(z)}{\diff z}$. Integrating by parts we obtain
\begin{multline}\label{intest}
  \int_0^{J\be/4}\diff y\,h'(y/\be J)\ln\left(\la_1[e^{-y}]\right)
  =\frac{\be J}4\ln\left(\la_1[e^{-J \be/4}]\right)\\
  +\int_0^{J\be/4}\diff y\,h(y/\be J)\,e^{-y}\,\frac{\la_1'[e^{-y}]}{\la_1[e^{-y}]},
\end{multline}
where
\[
  \la_1'[a]:=\frac{\diff\la_1[a]}{\diff a}.
\]
By Puiseux's theorem, $\la_1[a]$ can be represented by a series in fractional powers of $a$
convergent on a neighborhood of $a=0$. In fact, using the explicit form~\eqref{eigveq} of the
eigenvalue equation it is readily shown that the Puiseux series for $\la_1[a]$ centered at $a=0$
is actually a power series in $a^{1/m}$, namely
\begin{equation}\label{Puis}
  \la_1[a]=1+(2^{n/m}-\de_{1m})a^{1/m}+\sum_{k=2}^\infty c_ka^{k/m}.
\end{equation}
Hence the first term in Eq.~\eqref{intest} is $O(\be e^{-J\be/4m})$ as $\be\to\infty$. On the
other hand, from the elementary inequalities
\[
  0\le h(z)\le 4z^2
\]
for $0\le z\le1/4$ it follows that
\begin{multline*}
  J\be\left|\int_0^{J\be/4}\diff y\,h(y/\be J)\,e^{-y}\,\frac{\la_1'[e^{-y}]}{\la_1[e^{-y}]}\right|\\
  \le\frac{4T}J\int_0^{J\be/4}\diff y\,y^2e^{-y}\left|\frac{\la_1'[e^{-y}]}{\la_1[e^{-y}]}\right|,
\end{multline*}
which tends to $0$ as $T\to0+$ since the integral
\[
\int_0^{\infty}\diff y\,y^2e^{-y}\left|\frac{\la_1'[e^{-y}]}{\la_1[e^{-y}]}\right|
\]
is convergent (indeed,
\[
  y^2e^{-y}\frac{\la_1'[e^{-y}]}{\la_1[e^{-y}]}=O\left(y^2e^{-y/m}\right)
\]
as $y\to\infty$ on account of Eq.~\eqref{Puis} ). Thus the integral in Eq.~\eqref{asyint} is
indeed $o(1)$ as $\be\to\infty$, which completes the proof of Eq.~\eqref{asyrel}.

\section{The dilogarithm function}\label{app.dilog}
In this section we shall collect a few well-known results about the dilogarithm function used in
the computation of the central charge $c$ in Section~\ref{sec.CB}. By definition, the dilogarithm
function $\Li_2(z)$ is given by
\[
  \Li_2(z)=-\int_0^z\frac{\ln(1-t)}t\,\diff t,
\]
where the integral can be taken along any path in the complex $t$ plane joining the origin to the
point $z$ and not intersecting the branch cut $[1,\infty)$ of $\ln(1-t)$. (Here and in what
follows $\ln$ denotes the principal branch of the logarithm, with imaginary part in $(-\pi,\pi]$.)

To begin with, from the definition of $\Li_2$ and the power series
\[
  -\ln(1-t)=\sum_{n=1}^\infty\frac{t^n}n
\]
it easily follows that
\begin{equation}\label{Li21}
  \Li_2(1)=\sum_{n=1}^\infty\frac1{n^2}=\ze(2)=\frac{\pi^2}6.
\end{equation}
We shall next prove the identity
\[
  \Li_2(z)+\Li_2(1-z)=\Li_2(1)-\ln z\ln(1-z).
\]
To this end, note that
\begin{align*}
  \Li_2(1-z)&=-\int_0^{1-z}\frac{\ln(1-t)}t\,\diff t=
              \int_1^z\frac{\ln t}{1-t}\,\diff t\\
            &=\Li_2(1)
              +\int_0^z\frac{\ln t}{1-t}\,\diff t,
\end{align*}
and hence
\begin{align*}
  &\Li_2(z)+\Li_2(1-z)=\Li_2(1)+\int_0^z\diff t\left(\frac{\ln t}{1-t}-\frac{\ln(1-t)}t\right)\\
  &=\Li_2(1)-\int_0^z\diff t\,\frac{\diff}{\diff t}\,\Big(\ln t\ln(1-t)\Big)\\
    &=\Li_2(1)-\ln z\ln(1-z),
\end{align*}
as claimed. From the last equality it immediately follows that
\begin{equation}\label{Li212}
  \Li_2(1/2)=\frac12\Li_2(1)-\frac12\ln^2(1/2)=\frac{\pi^2}{12}-\frac12\ln^22,
\end{equation}
which can be used to evaluate the following integral:
\begin{align*}
  \int_0^1\frac{\ln(1-t)}{2-t}\,\diff t
  &=\int_1^2\frac{\ln(t-1)}{t}\,\diff t\\
  &= \int_1^2\frac{\ln t}t\,\diff t+\int_1^2\frac{\ln(1-t^{-1})}t\,\diff t\\
  &=\frac12\ln^22+\int^1_{1/2}\frac{\ln(1-t)}t\,\diff t\\
  &=\frac12\ln^22-\Li_2(1)+\Li_2(1/2)=-\frac{\pi^2}{12}.
\end{align*}
Finally, we shall compute $\Li_2(-1)$, which was implicitly used to evaluate the
partition function for the $\su(k|k)$ PF chain (cf.~Eq.~\eqref{fkkPF}):
\begin{align*}
  \Li_2(-1)&=-\int_0^{-1}\diff t\,\frac{\ln(1-t)}t=\sum_{n=1}^\infty\frac{(-1)^n}{n^2}\\
           &=-\sum_{k=0}^\infty\frac{1}{(2k+1)^2}+\sum_{k=1}^\infty\frac{1}{4k^2}
             =-\frac12\sum_{k=1}^\infty\frac{1}{k^2}\\
  &=-\frac{\pi^2}{12}.
\end{align*}


%

\end{document}